\documentclass[twocolumn]{aa} % for a paper on 2 columns
\usepackage{graphicx}
\usepackage{txfonts}
\usepackage{subfigure}
\usepackage{natbib}
\usepackage[hidelinks]{hyperref}

\newcommand{\Ec}{E_\mathrm{C}}

\begin{document}

   \title{Chromospheric emission from nanoflare heating in RADYN simulations}
   
%   \subtitle{Subtitle}

   \author{H.~Bakke\inst{\ref{inst1}, \ref{inst2}}
          \and
          M.~Carlsson\inst{\ref{inst1}, \ref{inst2}}
          \and
          L.~Rouppe van der Voort\inst{\ref{inst1}, \ref{inst2}}
          \and
          B.~V.~Gudiksen\inst{\ref{inst1}, \ref{inst2}}
          \and
          V.~Polito\inst{\ref{inst3}, \ref{inst4}}
          \and
          P.~Testa\inst{\ref{inst5}}
          \and
          B.~De~Pontieu\inst{\ref{inst4}, \ref{inst1}, \ref{inst2}}
          % Mats, Luc, Boris, Vanessa, Paola, Bart
          }

   \institute{
   Institute of Theoretical Astrophysics, 
   University of Oslo,
   P.O.Box 1029 Blindern, 
   N-0315 Oslo,
   Norway \label{inst1}
   \and
   Rosseland Centre for Solar physics,
   University of Oslo, 
   P.O.Box 1029 Blindern, 
   N-0315 Oslo,
   Norway \label{inst2}
   \and
   Bay Area Environmental Research Institute,
   NASA Research Park,
   Moffett Field,
   CA 94035,
   USA \label{inst3}
   \and
   Lockheed Martin Solar \& Astrophysics Laboratory,
   3251 Hanover St,
   Palo Alto,
   CA 94304,
   USA \label{inst4}
   \and
   Harvard-Smithsonian Center for Astrophysics,
   60 Garden St,
   Cambridge,
   MA 02193,
   USA \label{inst5}
   }

   \date{}

% \abstract{}{}{}{}{} 
% 5 {} token are mandatory
 
  \abstract
   {Heating signatures from small-scale magnetic reconnection events in the solar atmosphere have proven to be difficult to detect through observations. Numerical models that reproduce flaring conditions are essential in understanding how nanoflares may act as a heating mechanism of the corona.} 
   {We study the effects of non-thermal electrons in synthetic spectra from 1D hydrodynamic RADYN simulations of nanoflare heated loops to investigate the diagnostic potential of chromospheric emission from small-scale events.}
   {The \ion{Mg}{ii}~h and k, \ion{Ca}{ii}~H and K, \ion{Ca}{ii}~854.2~nm, and H$\alpha$ and H$\beta$ chromospheric lines were synthesised from various RADYN models of coronal loops subject to electron beams of nanoflare energies. The contribution function to the line intensity was computed to better understand how the atmospheric response to the non-thermal electrons affects the formation of spectral lines and the detailed shape of their spectral profiles.}
   {The spectral line signatures arising from the electron beams highly depend on the density of the loop and the lower cutoff energy of the electrons. Low-energy (5~keV) electrons deposit their energy in the corona and transition region, producing strong plasma flows that cause both redshifts and blueshifts of the chromospheric spectra. Higher-energy (10 and 15~keV) electrons deposit their energy in the lower transition region and chromosphere, resulting in increased emission from local heating. Our results indicate that effects from small-scale events can be observed with ground-based telescopes, expanding the list of possible diagnostics for the presence and properties of nanoflares.}
   {}


   \maketitle
%
%________________________________________________________________

\section{Introduction}

Nanoflares are small-scale events associated with magnetic reconnection in the solar atmosphere. The nanoflare heating mechanism is one of the prime candidates in understanding the high temperature of the corona \citep{1988ApJ...330..474P}. % Parker nanoflare
Flare energy is believed to be transported by electrons from the thermal background, where the electrons are accelerated to non-thermal energies as the magnetic field lines reconnect. The accelerated electrons travel along the magnetic field through the transition region (TR) and chromosphere, where the energy is lost via Coloumb collisions with the ambient plasma \citep{1971SoPh...18..489B, 1978ApJ...224..241E, 2011SSRv..159..107H}. % Brown deduction; Emslie collisional; Holman+ implications
This gives rise to observable signatures in spectral lines that are formed in the sites where energy is deposited. However, spectral line signatures are also affected by the response to the dissipated electron energy elsewhere in the atmosphere, such as strong plasma flows. 

Signatures of non-thermal electrons are found in spectra from hard X-ray observations of active region flares. 
However, X-ray observations of small-scale events with energies in the range $10^{24}$--$10^{25}$~erg remain rare because signatures from non-thermal electrons are typically below the detection threshold \citep[although, see e.g.][]{2017ApJ...844..132W, 2020ApJ...891L..34G, 2021MNRAS.507.3936C}. % Wright+ microflare; Glesener+ accelerated; Cooper+ NuSTAR
Indications of energetic events are obtained in regions that are responsive to heating. In the corona, such indications are more difficult to observe because its high conductivity smears the heating signatures out. 
The TR, on the other hand, is highly responsive to heating because of rapid changes in density, temperature, and volume during heating events. Non-thermal electrons from coronal heating events give rise to changes in density and temperature through collisions with the dense TR and chromospheric plasma.
Through advanced numerical simulations, \citet{2014Sci...346B.315T} % Testa+ evidence
have demonstrated that blueshifts in the \ion{Si}{iv}~140.2~nm line observed with the Interface Region Imaging Spectrograph \citep[IRIS;][]{2014SoPh..289.2733D} % De Pontieu+ IRIS 
in small heating events at the footpoints of transient hot loops are consistent with heating by non-thermal electrons, and they cannot be reproduced by models that assume heating only by thermal conduction. \citet{2018ApJ...856..178P} % Polito+ response 
carried out an extensive numerical investigation to better understand and interpret TR observations by exploring a wide region of parameters. 
\citet{2020ApJ...889..124T} % Testa+ variability 
have further shown that observations of high variability ($\lesssim 60$~s) at the footpoints of active region (AR) high-temperature coronal loops ($\sim$ 8-10~MK) in combination with numerical simulations provide powerful diagnostics of coronal heating and energy transport.

\begin{figure}[!ht] 
    \includegraphics[width=\columnwidth]{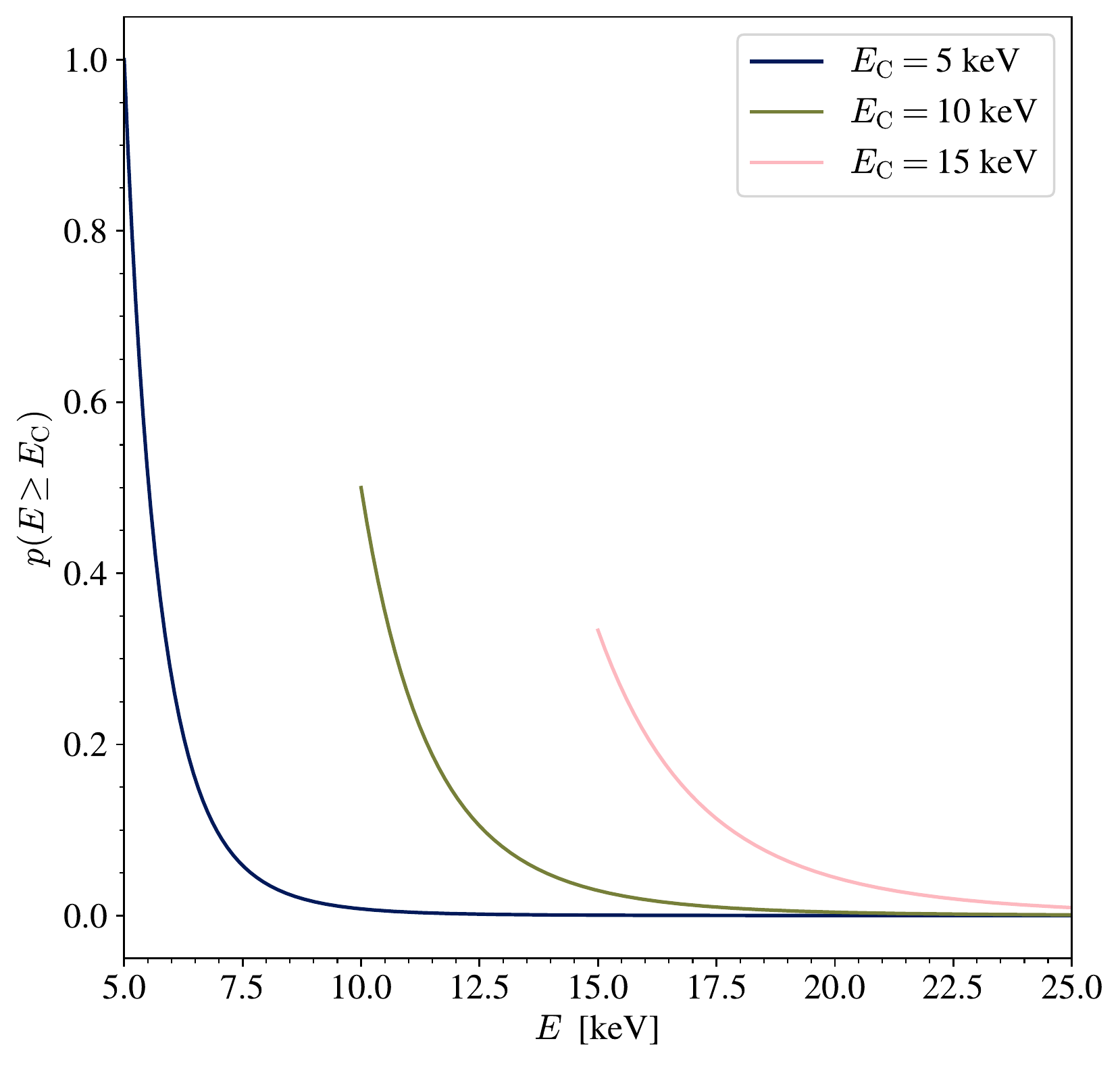}
    \centering
    \caption{Sketch of power-law distributions with different cutoff energies $\Ec$, on a linear scaling. The figure serves as a visual aid to distinguish the difference between the electron distributions.}
    \label{fig:e_dist}
\end{figure}

\citet{2018ApJ...856..178P} % Polito+ response 
analysed in detail the TR \ion{Si}{iv} lines and upper chromospheric \ion{Mg}{ii}~h and k lines from RADYN 1D simulations of nanoflare heated loops, and \citet{2020ApJ...889..124T} also investigated \ion{C}{ii} and \ion{Mg}{ii} triplet emission from the same set of simulations. In this work, we extend their analyses by probing the impact of electron beam heating on spectral lines that are formed deeper in the atmosphere and are readily accessible by ground-based telescopes: \ion{Ca}{ii}~H and K, \ion{Ca}{ii}~854.2~nm, and H$\alpha$ and H$\beta$. 
We focus our efforts on the electron beam heating models with the shorter loop length (15~Mm) and initial loop-top temperatures of 1 and 3~MK, representing respectively an initially almost empty loop and a typical AR core loop. 
The selection of simulations is motivated by the study of nanoflare electron beams in 3D MHD Bifrost models \citep{2018A&A...620L...5B, 2020A&A...643A..27F}, % Bakke+ non-thermal, Frogner+ accelerated
which only covers cooler and shorter loops. With RADYN we can more easily study the details of such conditions, while with Bifrost this is more difficult due to the complex configuration of the simulation.
We reproduce the upper chromospheric \ion{Mg}{ii}~h results of \citet{2018ApJ...856..178P} % Polito+ response 
with the aim of making a direct comparison with the lower chromospheric observables.
In the analysis we explore the impact of varying the low-energy cutoff value by investigating the atmospheric response, the Doppler shift of the chromospheric lines, and the formation of line intensity.

\section{Method}

\subsection{The RADYN numerical code}

The numerical simulations were performed using the RADYN numerical code \citep{1992ApJ...397L..59C, 1995ApJ...440L..29C, 1997ApJ...481..500C, 2015TESS....130207A}. % Carlsson&Stein RADYN; Allred+ RADYN flare
RADYN solves the equation of charge and population conservation coupled to the non-linear equations of hydrodynamics in a 1D plane-parallel atmosphere on an adaptive grid \citep{1987JCoPh..69..175D}. % Dorfi&Drury grids
The non-local thermodynamic equilibrium (non-LTE) radiation transport is solved for H, He, and \ion{Ca}{ii}, while continua from other atomic species are treated in local thermodynamic equilibrium (LTE) as background metal opacities using the Uppsala opacity package \citep{gustafsson1973}. % Gustafsson fortran
Optically thin losses are calculated using the CHIANTI 7.1 atomic database \citep{1997A&AS..125..149D}. % Dere+ CHIANTI

The RADYN version described in \cite{2015TESS....130207A} % Allred+ RADYN flare
allows to model the effect of non-thermal electrons in flaring loops.
The particle distribution function is calculated by the Fokker-Planck equations, and the electron beam spectrum follows a power-law. 
With RADYN flare models it is possible to study the atmospheric response to energy deposited in the loop. 
RADYN models half of the loop only, between one footpoint and the loop apex. It is implicitly assumed that the full loop is symmetric around the apex.
Further, the half-loop is modelled as a quarter circle, where the geometry is used to vary the value for the gravitational acceleration along the $z$-axis. The radiative transfer is computed by assuming a 1D atmosphere along the $z$-axis. The spectral diagnostics of interest are formed in the lowest part of the atmosphere where the loop geometry is near-vertical. This allows for comparisons with observations with a top-down line of sight.

We have investigated the heating models of coronal nanoflares carried out by \cite{2018ApJ...856..178P}, % Polito+ response 
but focused our efforts on the models with transport by non-thermal electrons. The atmospheric structure included a plage-like chromosphere following the work by \citet{2015ApJ...809L..30C}. % Carlsson+ plage
The numerical investigation included several nanoflare models with different loop-top temperatures and parameters of the non-thermal electron power-law distribution.

\subsection{Parameter survey}

A total of six numerical simulations were analysed. The energy deposited in a single loop was set to $E = 6 \cdot 10^{24}$~erg in all models, following the work by \citet{2014Sci...346B.315T, 2020ApJ...889..124T}. % Testa+ evidence; Testa+ variability
The cross-sectional area of the loop was set to $A = 5 \cdot 10^{10}$~m$^2$, corresponding to a diameter of about 250~km. 
The half-loop lengths were set to 15~Mm with initial loop-top temperatures $T_\mathrm{LT} = 1$ and 3~MK, giving two initial atmospheres that aimed to reproduce AR loops at different heating stages. With the assumption of an atmosphere in hydrostatic equilibrium, low-temperature loops have low density while hotter loops are also denser. The 1~MK loop represents an empty strand where heating has not yet occurred, and has an apex electron number density around $10^{8.7}$~cm$^{-3}$. The 3~MK loop represents a previously heated strand with apex electron number density of approximately $10^{9.6}$~cm$^{-3}$. In the following, we refer to the 1~MK and 3~MK loops as empty and pre-heated loops, respectively.
 
In RADYN flare models, the energy flux $F$, spectral index $\delta$, and low-energy cutoff $\Ec$ that go into the power-law distribution are user-specified. The input value for the electron beam energy flux was calculated as
\begin{equation}
    F = \frac{E}{At} = 1.2 \cdot 10^9~\mathrm{erg}~\mathrm{cm}^{-2}~\mathrm{s}^{-1},  
\end{equation}
where $E$ and $A$ are the total energy and cross-section of the loop given above. The loop was heated constantly for $t = 10$~s which is consistent with IRIS observations of TR moss showing that the lifetime of short-lived footpoint brightenings varies between 10 -- 30~s \citep{2013ApJ...770L...1T, 2014Sci...346B.315T, 2020ApJ...889..124T}. % Testa+ observing; Testa+ evidence; Testa+ variability

A large spectral index ($\delta = 7$) was chosen for the energy distribution, supported by observational evidence of increasing spectral index with decreasing flare energy \citep{2011SSRv..159..263H}, % Hannah+ microflares
and compatible with values from rare hard X-ray observations of these small heating events \citep[e.g.][]{2020ApJ...891L..34G, 2021MNRAS.507.3936C}. % Glesener+ accelerated; Cooper+ NuSTAR
The low-energy cutoff values $\Ec = 5$, 10, and 15~keV were chosen to represent beams that are dominated by low-, intermediate-, and high-energy electrons. Figure~\ref{fig:e_dist} shows examples of power-law distributions for the different values of $\Ec$, where the total energy of the beam is fixed. Each distribution starts at its respective cutoff energy (which is the energy that most of the electrons have because of the large spectral index), before exponentially declining as the energy increases. The $\Ec = 5$~keV distribution shows that there are few electrons with $E > 9$~keV, and that most of the electrons have energies between 5 and 6~keV. As the cutoff energy increases, the number of electrons with $E = \Ec$ decreases. However, there are generally more electrons with higher energies in these distributions.

\begin{figure*}[!ht]
    \centering
    \includegraphics[width=\textwidth]{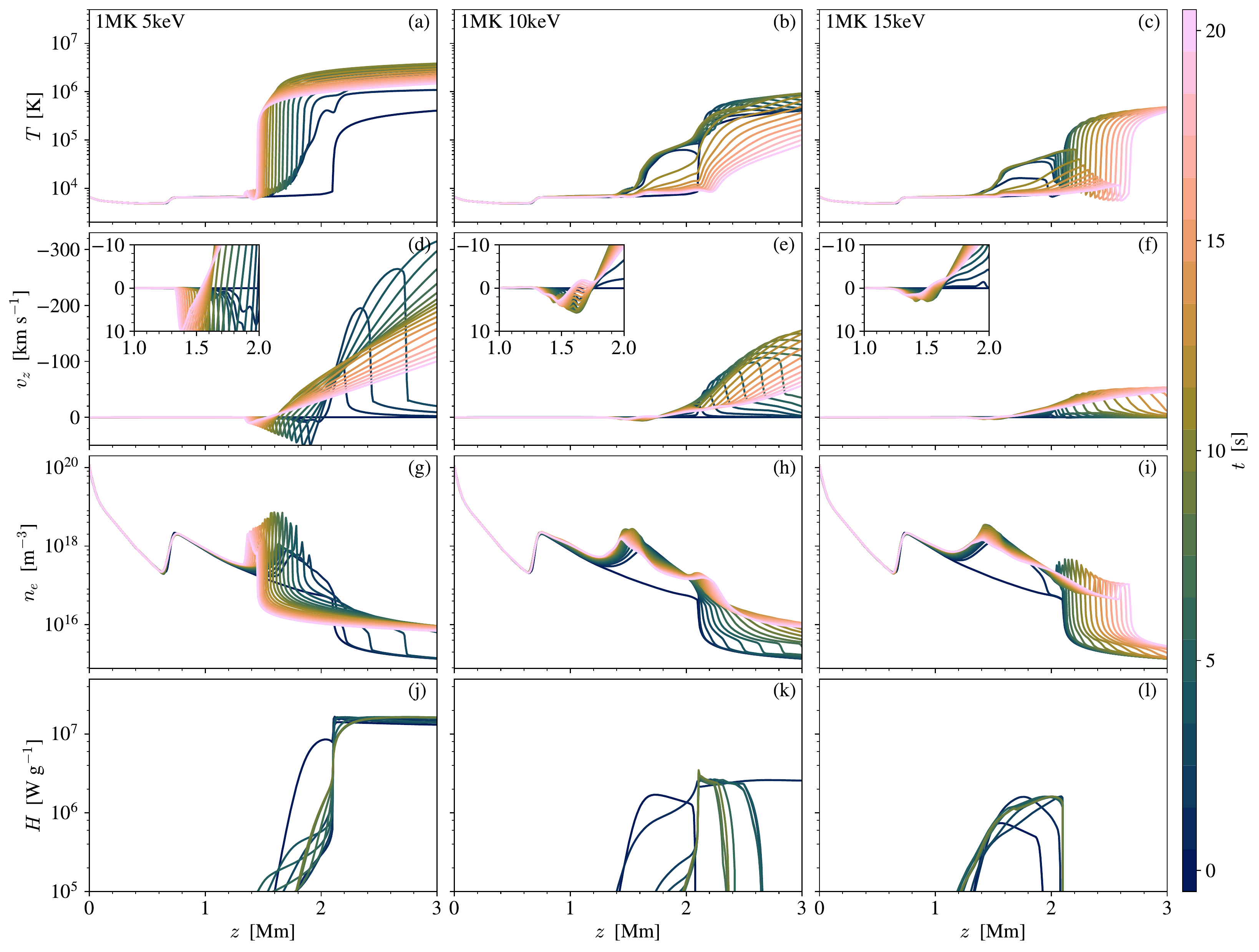}
    \caption{Atmospheric response to the RADYN simulation for the 15~Mm half-loop with initial apex temperature of 1~MK. Each column corresponds to the heating model with $\Ec = 5$, 10, and 15~keV, and each row represents the temperature, velocity, electron number density, and beam heating rate. Negative (positive) velocities correspond to upflows (downflows). The quantities are plotted in the range $z \in [0,3]$~Mm at 1~s intervals. Temperature, velocity, and electron number density are displayed for the first 20~s of the simulations, while the beam heating rate is given for the duration of the injected electron beam (10~s). The insets in panels (d)-(f) show the velocity in the region $z = $~1--2~Mm saturated to $\pm 10$~km~s$^{-1}$.}
    \label{fig:1MK_ar}
\end{figure*}

\begin{figure*}[!ht]
    \centering
    \includegraphics[width=\textwidth]{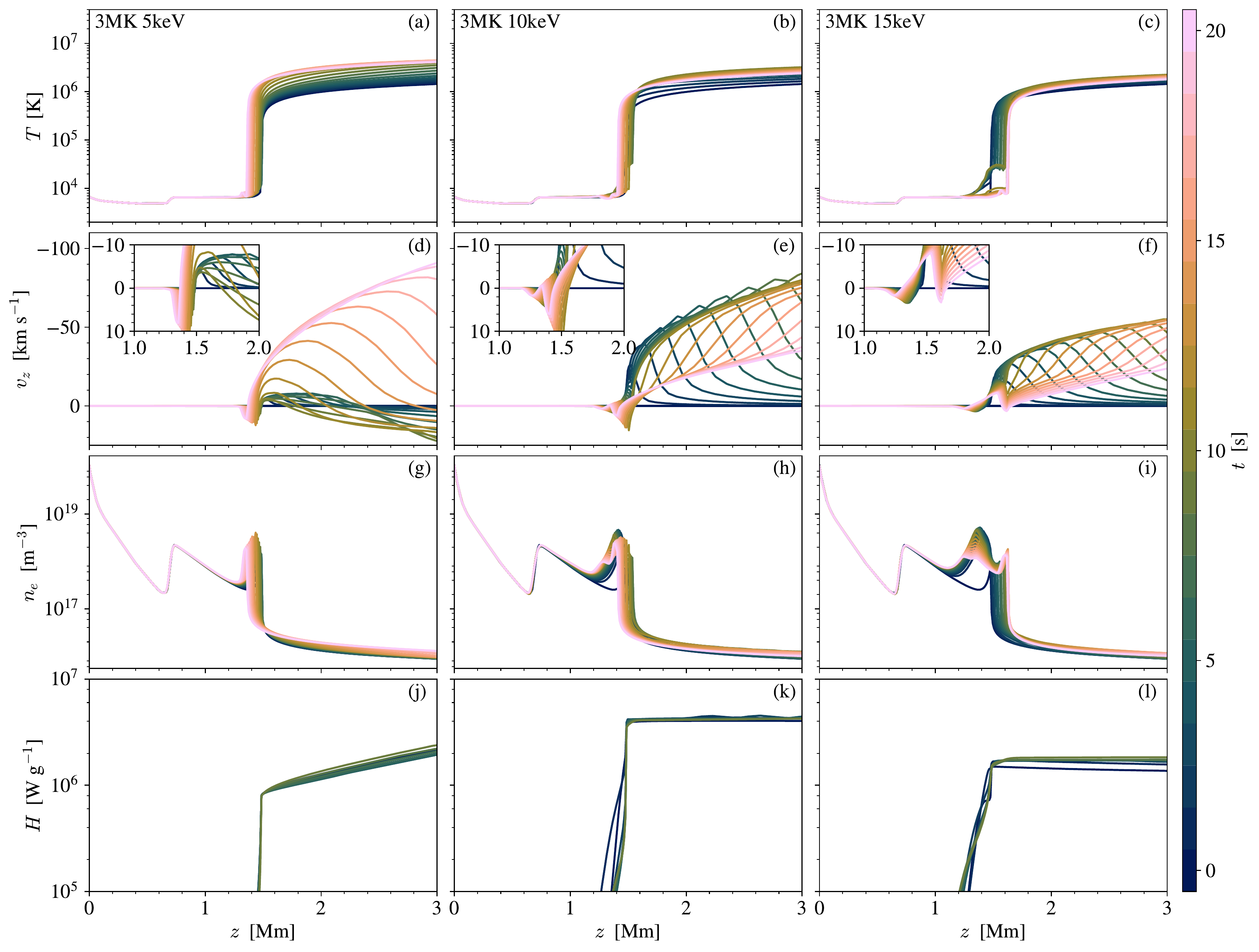}
    \caption{Atmospheric response to the RADYN simulation for the 15~Mm half-loop with initial apex temperature of 3~MK. See Fig.~\ref{fig:1MK_ar} for an explanation of the different panels.}
    \label{fig:3MK_ar}
\end{figure*}

\subsection{Spectral synthesis with RH}

The chromospheric lines were synthesised using the RH1.5D radiative transfer code \citep{2001ApJ...557..389U, 2015A&A...574A...3P}. % Uitenbroek RH, Pereira & Uitenbroek RH1.5D
RH1.5D solves the non-LTE radiative transport for spectral lines in partial redistribution (PRD). This is important in the synthesis of spectral lines where there is a correlation between the frequencies of the absorbed and re-emitted radiation such that the line source function varies with frequency. RADYN assumes complete redistribution, where the frequencies of the ingoing and outgoing photons are independent. While this is reasonable for the energetics in RADYN simulations, the assumption is best suited for modelling of photospheric lines. A more accurate treatment of photon scattering is required when modelling chromospheric lines, hence PRD is generally assumed in the synthesis of \ion{Mg}{ii}~h and k \citep{1974ApJ...192..769M, 2013ApJ...772...89L, 2013ApJ...772...90L} % Milkey&Mihalas PRD Mg II, Leenarts+ IRIS diagnostics (a), Leenarts+ IRIS diagnostics (b)
and \ion{Ca}{ii}~H and K \citep{1974SoPh...38..367V, 1975ApJ...199..724S, 2018A&A...611A..62B}. % Vardavas&Cram partially, Shine+ PRD Ca II, Bjørgen+ 3D

The atmospheric evolution from the RADYN simulations was used as input to the RH code at every second. We added a microturbulence term of 7~km~s$^{-1}$ to produce a plage-like chromosphere following the work of \citet{2015ApJ...809L..30C}. % Carlsson+ plage
This atmospheric structure is not proposed as a realistic model of plage, but aims to reproduce the observed average \ion{Mg}{ii}~k line profile. However, we note that the atmosphere still fails to reproduce the single peaked profiles observed in plage. 
 
In the synthesis of hydrogen and calcium, we used the 5 level-plus-continuum \ion{H}{i} and \ion{Ca}{ii} atoms that are standard in RH1.5D, while for magnesium, we used the 10 level-plus-continuum \ion{Mg}{ii} atom from \citet{2013ApJ...772...89L}. % Leenaarts+ formation I
All lines were treated in PRD, even though \citet{2018A&A...611A..62B} % Bjørgen+ 3D
found that it is less important for \ion{Ca}{ii}~854.2~nm. The synthetic spectra were calculated for the first 40~s, allowing us to investigate the chromospheric emission during the heating and post-heating phases in detail.

\citet{2020ApJ...889..124T} % Testa+ variability
investigated the \ion{Mg}{ii}~279.882~nm triplet in a variety of RADYN simulations, including the empty loop models presented here. The \ion{Mg}{ii} triplet lines (279.160~nm, 279.875~nm and 279.882~nm) are located in the wings of the h and k lines, and are expected to be in emission when large temperature increases are present in the lower chromosphere \citep{2015ApJ...806...14P}. % Pereira+ triplet
We included the \ion{Mg}{ii}~279.882~nm triplet in the formation height analysis, in order to compare it to the \ion{Ca}{ii}~854.2~nm line which is formed in the lower chromosphere.

\subsection{Contribution function to the line intensity}

The final part of the spectral diagnostics consisted of analysing the contribution function to the emergent intensity. Following \citet{1997ApJ...481..500C}, % Carlsson&Stein formation
the contribution function was calculated as
\begin{equation} \label{eq:cont_func}
    C_{I_\nu}(z) \equiv \frac{\mathrm{d}I_\nu(z)}{\mathrm{d}z} = S_\nu \ \tau_\nu \mathrm{e}^{-\tau_\nu} \ \frac{\chi_\nu}{\tau_\nu}.    
\end{equation}
The first term on the right-hand side gives the total source function $S_\nu$, which is dependent on frequency because we assume PRD. The optical depth factor $\tau_\nu \mathrm{e}^{-\tau_\nu}$ represents the Eddington-Barbier part of the contribution function, and peaks at $\tau_\nu = 1$. The $\chi_\nu/\tau_\nu$ term is the ratio of the opacity over optical depth which is sensitive to velocity gradients in the atmosphere, and is responsible for the line asymmetries. The factor is dominant when the opacity is large at small optical depths, which typically occurs when the velocity gradients are strong. 

The opacity $\chi_\nu$ and source function $S_\nu$ were calculated by the RH1.5D code for selected wavelengths, and the optical depth $\tau_\nu$ was calculated by integrating the opacity over the entire loop length. The contribution function to the line intensity was calculated in the range $z = $~0--2~Mm, which is the part of the loop that has near-vertical geometry.

\section{Results}

\begin{figure*}[!ht]
    \centering
    \includegraphics[width=18cm]{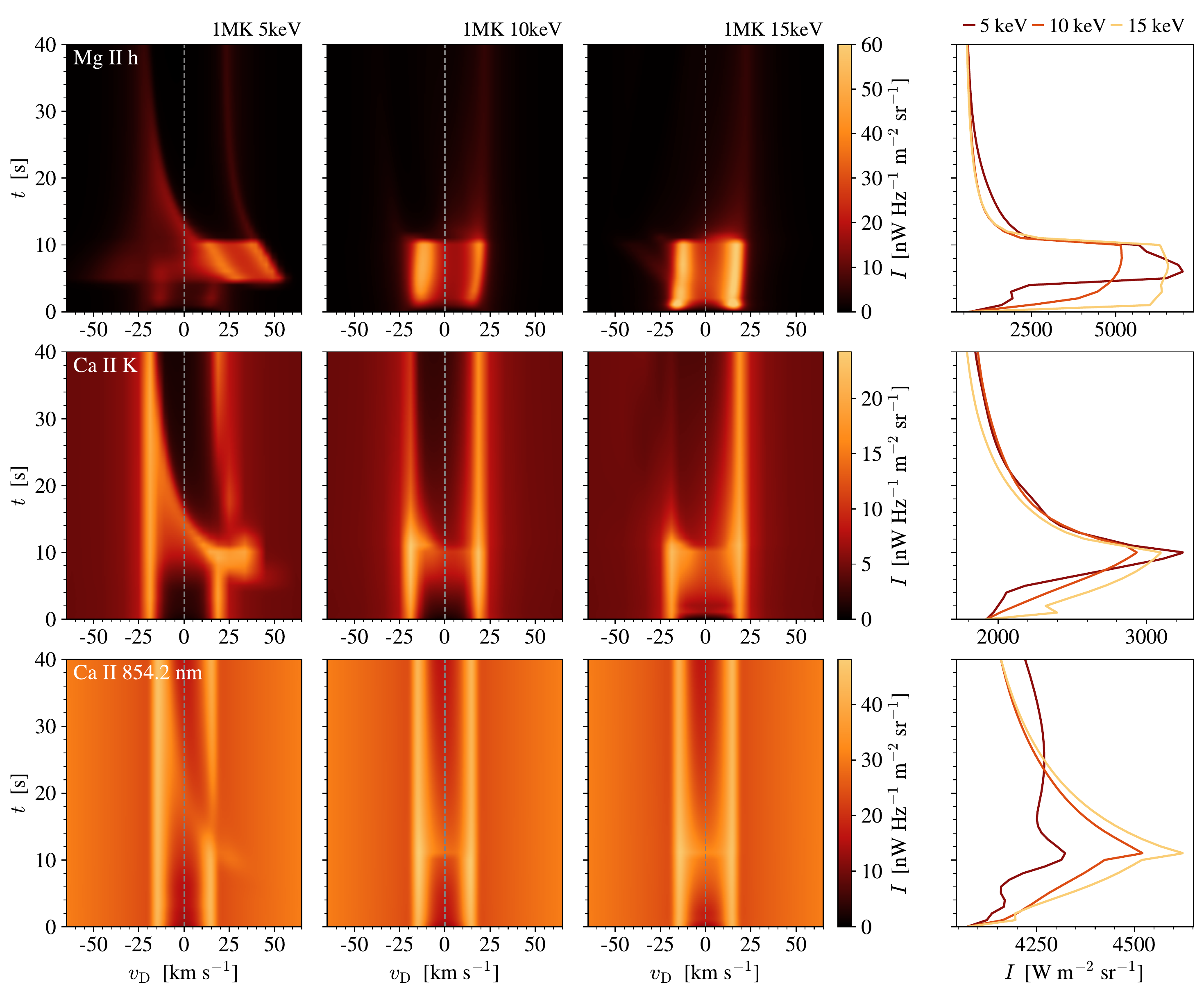}
    \caption{Spectral evolution of \ion{Mg}{ii}~h, \ion{Ca}{ii}~K, and \ion{Ca}{ii}~854.2~nm spectra for the heating models with $T_\mathrm{LT} = 1$~MK and $\Ec = 5$, 10, and 15~keV. The $x$-axes in the first three panels of each row are in units of Doppler offset, where negative (positive) velocities indicate blueshifts (redshifts). The panels of the right column show light curves from each heating model, where the $x$-axes are in units of integrated intensity. The first 40~s of the simulations are shown, including the 10~s electron injection phase. The \ion{Mg}{ii}~h spectra are clipped at 60~nW~Hz$^{-1}$~m$^{-2}$~sr$^{-1}$ to emphasise the less bright features. Line profiles for $t = 0$, 7, and 15~s are shown in Fig.~\ref{fig:mimgca}.}
    \label{fig:1MK_hm}
\end{figure*}

\begin{figure*}[!ht]
    \centering
    \includegraphics[width=18cm]{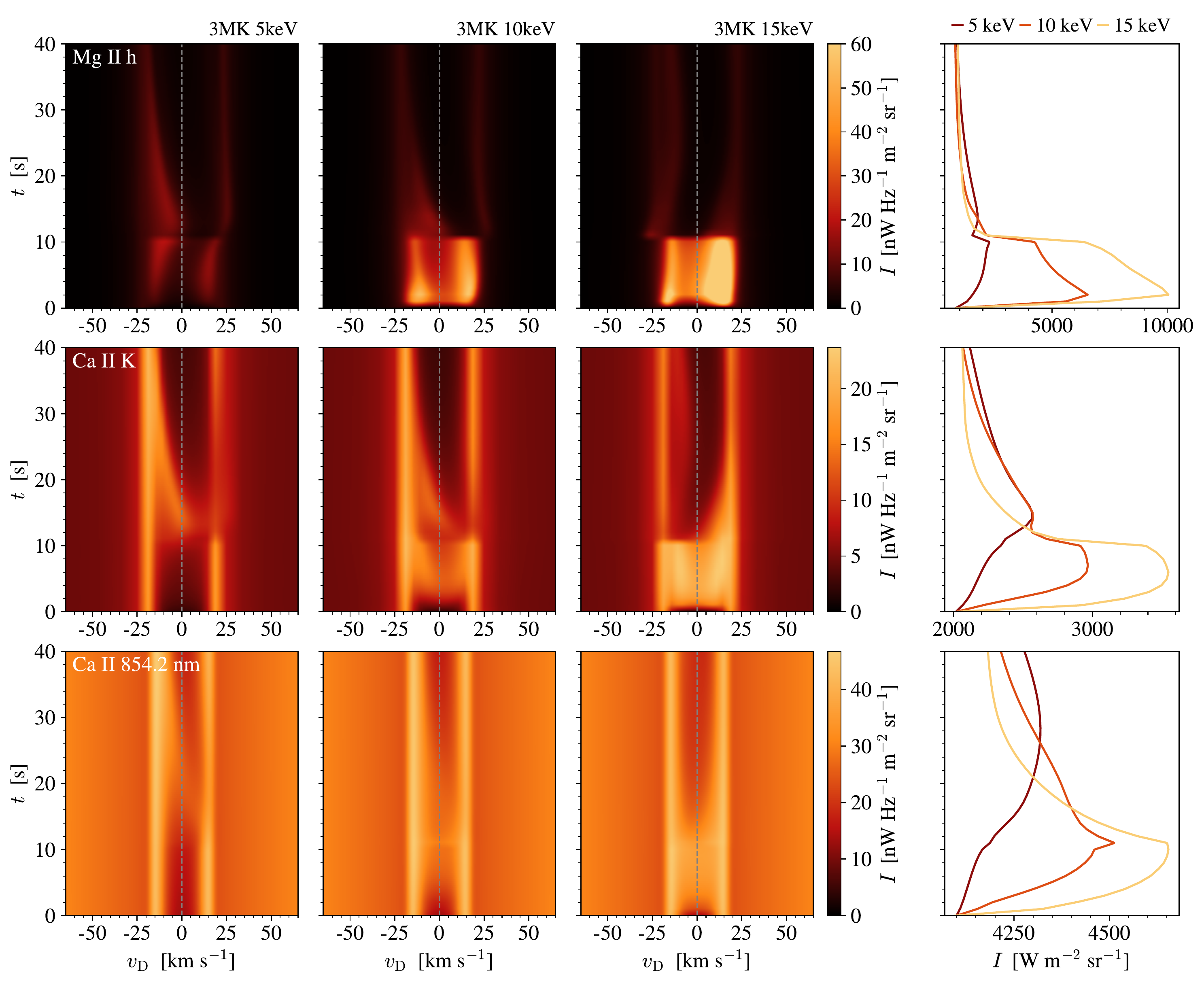}
    \caption{Synthetic spectra for the $T_\mathrm{LT} = 3$~MK heating models. See the caption of Fig.~\ref{fig:1MK_hm} for more details.}
    \label{fig:3MK_hm}
\end{figure*}

\subsection{Atmospheric response to non-thermal electrons}

Figures \ref{fig:1MK_ar} and \ref{fig:3MK_ar} show the evolution of temperature $T$, vertical velocity $v_z$, electron number density $n_e,$ and beam heating rate $H$ for the loops with $T_\mathrm{LT} = 1$ and 3~MK. The quantities are plotted in the range of interest in terms of formation heights for the relevant spectra. For the temperature, velocity, and density, we focus on the first 20~s of the simulations in order to investigate the heating and early relaxation phase in more detail. The beam heating rate is shown for the first 10~s, corresponding to the loop heating time. 
These figures essentially reproduce figures 2 and 3 of \citet{2018ApJ...856..178P}, % Polito+ response
but show the evolution in the lower 3~Mm of the loops in more detail. We refer to \citet{2018ApJ...856..178P} % Polito+ response
for more details on the evolution of the entire 15~Mm loop-length. 

\subsubsection{Empty loop}

The difference between heating models is largest in the loop with initial low electron number density and 1~MK loop-top temperature (see Fig.~\ref{fig:1MK_ar}).
During the 10~s heating phase, electrons with $\Ec = 5$~keV deposit most of their energy in the corona and TR, leading to an increase in temperature and density of the ambient plasma of about one order of magnitude. This causes the TR to shift to lower depths, as can be seen in panel (a). The large (negative) velocities in panel (d) result from an increase in pressure due to the rise in temperature. The upward motion of hot plasma ('chromospheric evaporation') increases with height, and as a consequence, the loops are filled with hot plasma and the electrons start depositing energy at increasing heights (see panel (j)). The positive velocities are indicative of a significant downflow of plasma towards the chromosphere, caused by the condensation of hot plasma filling the flaring loop. 

The heating models with $\Ec = 10$ and 15~keV (middle and right columns) have more energetic electrons in their distributions. Higher-energy electrons can travel through the corona depositing only a small fraction of their energy,
allowing them to penetrate deeper into the atmosphere. Panels (h) and (i) show that the deposited energy is at a maximum between $z \approx 1.5$ and 2.1~Mm within the first few seconds. The radiative losses from ionised hydrogen and helium are not able to balance the large amount of energy deposited, resulting in a temperature increase in this region (as can be seen in panels (b) and (c)). The upflowing velocities in panels (e) and (f) are lower compared to the 5~keV model. This is due to the difference in sites where electron energy is deposited. While the 5~keV electrons deposit most of their energy in the corona and TR, the 10 and 15~keV electrons mostly deposit their energy in the chromosphere where the plasma is denser. Dense plasma is harder to accelerate, and the combination of this and the fact that energy radiates away more efficiently contributes to weaker upflowing velocities. Consequently, the upflows cause the TR to move towards greater heights. 

The upward motion of hot plasma in the 10 and 15~keV heating models causes the density of the loops to increase over time, and similarly to the 5~keV model, the electrons start to deposit energy at increasing heights (see panels (k) and (l)). However, we note that in the 15~keV model the increase in height where energy is deposited does not change much over the 10~s heating duration compared to the other models. The 5~keV electrons are more effective at heating the corona, and the large upflowing velocity demonstrates that this heating model is more efficient at filling the flaring loop with high-temperature plasma. 

\subsubsection{Pre-heated loop}

Figure \ref{fig:3MK_ar} shows the atmospheric response to the heating models for the loops with initial apex temperature of 3~MK. As mentioned earlier, the hotter and denser loops represent strands that have been previously heated. Therefore, the initial atmospheres have transition regions that are located at lower depths compared to the 1~MK loop models. Panels (a)-(c) show a temperature increase in the corona for all $\Ec$ values. Due to the higher initial density of the loops, the injected electrons deposit more of their energy in the corona. In the 5~keV heating model, the energy deposited directly into the corona (as indicated by the increasing heating rate with height in panel (j), but also shown in \cite{2018ApJ...856..178P}) % Polito+ response
causes the temperature and pressure to increase, leading to both upflows and downflows of plasma. However, not all electrons deposit their energy in the corona. Some of the electrons are able to reach the TR, where they deposit their energy at $z \approx 1.5$~Mm. The local heating of the TR in combination with the conduction front from the corona results in a pressure gradient that produces both upward motion of plasma towards the corona as well as downflowing plasma into the TR. 

The intermediate- and high-energy electrons initially deposit their energy in the TR, resulting in upflows of hot plasma into the corona. As the loops are filled with high-temperature plasma, the electrons start to deposit energy at greater heights, as can be seen in panels (k) and (l). Panels (e) and (f) show that as the height where energy is deposited increases, the evaporation front travels upwards along the loop. The moving TR region in panel (c) shows resemblance to the 10 and 15~keV runs in Fig.~\ref{fig:1MK_ar}, where the properties of dense plasma in combination with direct heating results in an increased temperature. However, the atmospheric response is much slower for the 3~MK loop because the plasma is harder to accelerate, hence the TR does not move much during the first 20~s. 

\subsection{Chromospheric emission}

Figures~\ref{fig:1MK_hm} and \ref{fig:3MK_hm} represent the time evolution of the synthetic \ion{Mg}{ii}~h, \ion{Ca}{ii}~K, and \ion{Ca}{ii}~854.2~nm spectra, along with their light curves, from the different heating models. The light curves are obtained by integrating the emission over the spectra in the wavelength range for each timestep. We note that the \ion{Mg}{ii}~h results are similar to \ion{Mg}{ii}~k, and that the \ion{Ca}{ii}~K results are similar to \ion{Ca}{ii}~H. Therefore, we only present our findings from the \ion{Mg}{ii}~h and \ion{Ca}{ii}~K lines. \citet{2018ApJ...856..178P} % Polito+ response 
analysed the \ion{Mg}{ii}~h line in detail, and we reproduce the \ion{Mg}{ii}~h results in order to make a direct comparison with the \ion{Ca}{ii} lines and study the effect from energy deposited by non-thermal electrons. 

Figure~\ref{fig:1MK_hm} shows the chromospheric emission for the 1~MK loop, where each column represents the different cutoff energies $\Ec$. The three spectral lines in the $\Ec = 5$~keV model are behaving similarly, but with some variation. The \ion{Mg}{ii}~h and \ion{Ca}{ii}~K profiles have redshifted components that are caused by the downflow of cool plasma into the chromosphere, as seen in Fig.~\ref{fig:1MK_ar}~(d). The \ion{Mg}{ii}~h line also has a blueshifted component, and is characterised by multiple peaks that are Doppler shifted up to approximately $\pm 50$~km~s$^{-1}$ during the heating phase. The \ion{Ca}{ii}~K line shows multiple peaks in the redshifted component that are Doppler shifted to around $+40$~km~s$^{-1}$. The \ion{Ca}{ii}~854.2~nm line shows weak emission of the red wing at the end of the heating phase. The blue peak gets broader and has a higher intensity level during the post-heating phase, as a response to the upflow of plasma into the corona and TR. 

\begin{figure*}[!ht]
    \centering
    \includegraphics[width=\textwidth]{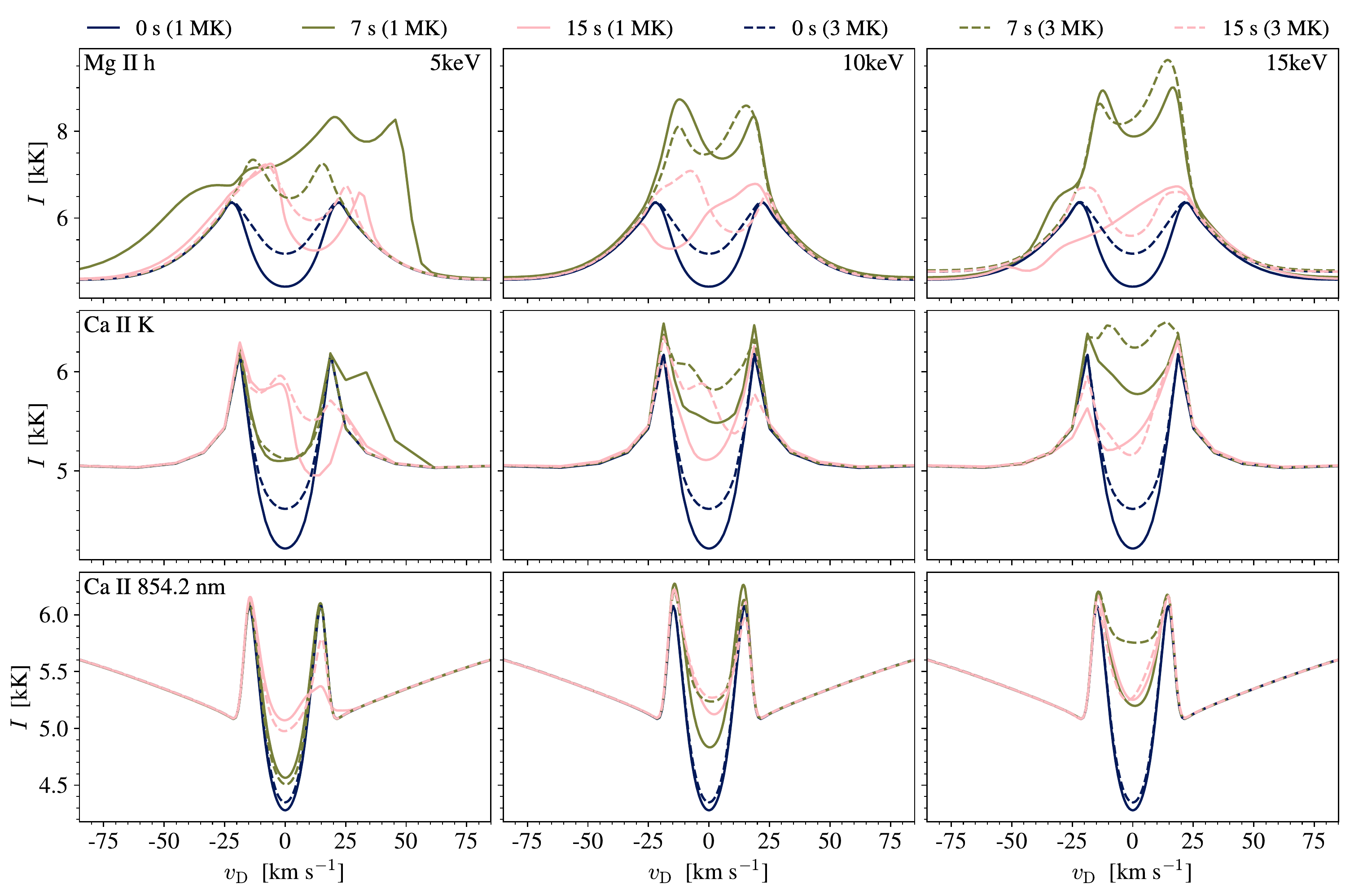}
    \caption{Synthetic \ion{Mg}{ii}~h, \ion{Ca}{ii}~K, and \ion{Ca}{ii}~854.2~nm line profiles. The solid and dotted lines represent the 1 and 3~MK loops, respectively. Line profiles during the pre-heating ($t = 0$~s), heating ($t = 7$~s), and post-heating ($t = 15$~s) phases are shown.}
    \label{fig:mimgca}
\end{figure*}

During the first 10~s of the simulations, while the electrons are still injected, the profiles in the $\Ec = 10$ and 15~keV models are mostly double peaked. The spectra experience an increase in line core intensity during the heating phase. This is because the high-energy electrons are able to penetrate deeper into the atmosphere before depositing their energy, as can be seen from panels (h) and (i) in Fig.~\ref{fig:1MK_ar}. During 5--10~s, the \ion{Mg}{ii}~h line in the 15~keV model shows faint emission of the blue wing, which is Doppler shifted to approximately $-30$~km~s$^{-1}$. After the heating phase, \ion{Mg}{ii}~h and \ion{Ca}{ii}~K are dominated by a red peak, and the profiles become almost single peaked. This is apparent from the solid pink lines in Fig.~\ref{fig:mimgca}.

The light curves from the \ion{Mg}{ii}~h spectra show a sudden increase in intensity as a response to the electron beams. The profile from the 5~keV model is generally less intense, but the integrated emission increases as a result of the broadening of the line. The light curves from the \ion{Ca}{ii} spectra have a slower increase and peak around 10~s. While most of the light curves decrease after the heating is turned off, the \ion{Ca}{ii}~854.2~nm integrated emission starts to increase around 15~s. This is due to the broadening and increased emission of the blue peak.

Figure~\ref{fig:3MK_hm} shows the spectra for the denser 3~MK loop. The 5~keV simulation shows qualitatively similar effects on the \ion{Mg}{ii}~h and \ion{Ca}{ii}~K lines, while the effect on the \ion{Ca}{ii}~854.2~nm line is less significant. The \ion{Mg}{ii}~h and \ion{Ca}{ii}~K lines both show broadening of the blue peak at around 10~s. During the post-heating phase, the entire \ion{Mg}{ii}~h line is shifted to the red before slowly reverting back. The blue component of \ion{Ca}{ii}~K is characterised by double peaks, and the general behaviour of the intensity after the heating is turned off is similar to that of \ion{Ca}{ii}~K in the 5~keV empty loop model. Due to the density of the loop, the electron energy is dissipated directly in the corona and TR. The chromosphere is not heated significantly, and the effects seen in \ion{Mg}{ii}~h and \ion{Ca}{ii}~K are due to local plasma flows.

In the 10 and 15~keV simulations, there is an increase in intensity as a function of $\Ec$ that is particularly apparent in \ion{Mg}{ii}. 
The peak intensity of \ion{Mg}{ii}~h in the 10~keV model is comparable to that of an empty loop. However, the energy deposited from the 15~keV electrons affect the chromospheric line greatly, where the intensity of the red peak is much stronger compared to the empty loop (see Fig.~\ref{fig:mimgca}). The \ion{Ca}{ii}~854.2~nm profiles in the $\Ec \geq 10$~keV simulations are comparable to the results of the empty loop, both in line features and evolution of the light curves. The \ion{Ca}{ii}~K profiles are greatly affected by the intermediate- and high-energy electrons, and are characterised by multiple peaks and strong emission of the line core. The latter is reflected in the light curves, which are similar to the \ion{Mg}{ii} integrated emission from the 10 and 15~keV models in that they have a steep decline. 

Figure~\ref{fig:mimgca} shows the \ion{Mg}{ii} and \ion{Ca}{ii} line profiles to better visualise the shapes and features that arise during the simulations. Particularly, the faint emission of the blue wing of \ion{Mg}{ii}~h in the 15~keV empty loop model is easier to observe (see the solid green line in the upper right panel). In the 5~keV empty loop model and the 5 and 10~keV dense loop models, the \ion{Ca}{ii}~K core is shifted to the red after the heating is turned off, and the blue component is double peaked.
The \ion{Ca}{ii}~854.2~nm profiles show that the response to the non-thermal electrons is mostly changing the absorption feature, which gets brighter as time progresses. In the 5~keV models, the red emission peak of \ion{Ca}{ii}~854.2~nm is weaker at later timesteps. 

\begin{figure*}[!ht]
    \centering
    \includegraphics[width=\textwidth]{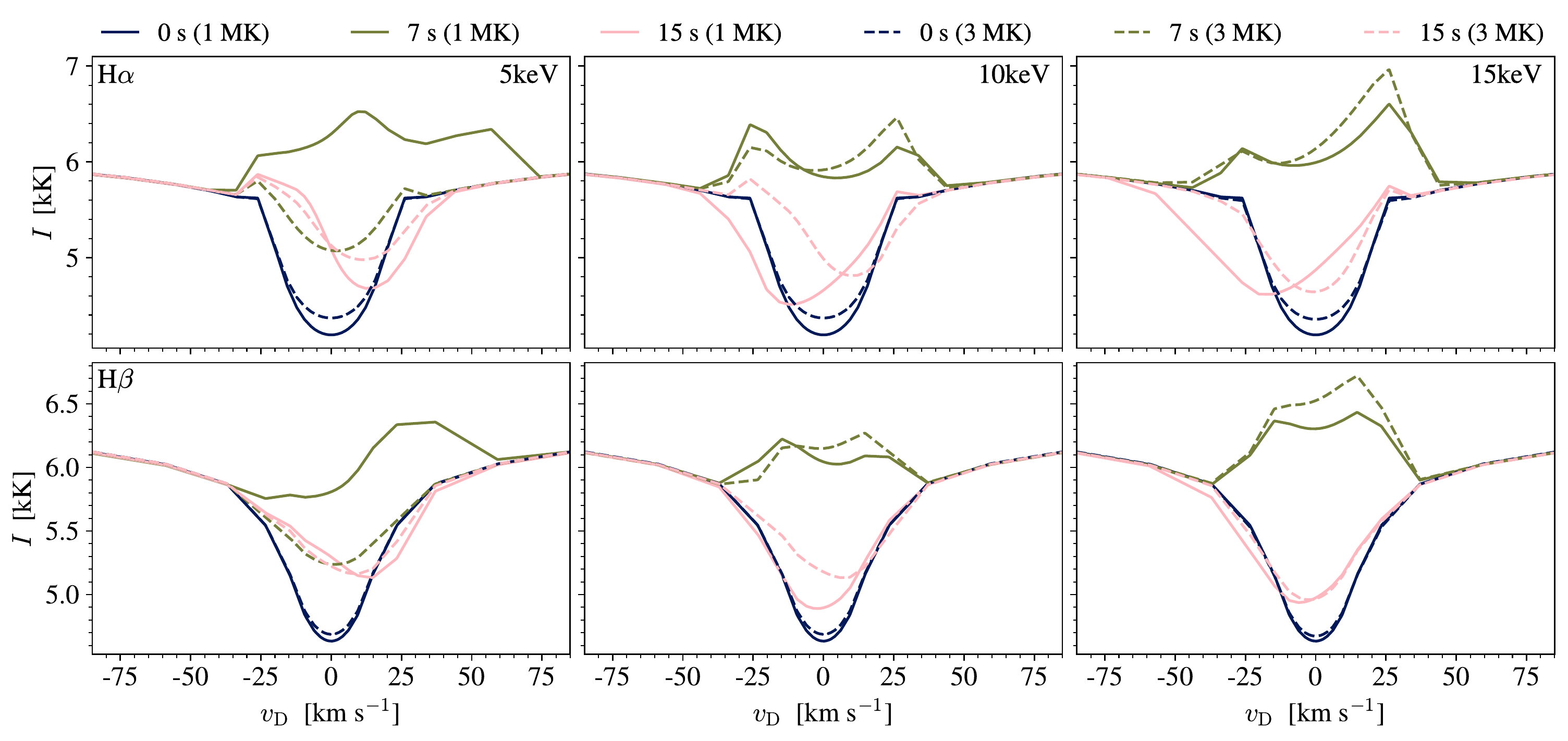}
    \caption{Synthetic H$\alpha$ and H$\beta$ spectra. See the caption of Fig.~\ref{fig:mimgca} for more details on the different panels.}
    \label{fig:mih}
\end{figure*}

Figure~\ref{fig:mih} represents the synthetic H$\alpha$ and H$\beta$ lines from all the simulations, and are given for the same timesteps as in Fig.~\ref{fig:mimgca}. 
The initial profiles are almost identical across the simulation span, where both H$\alpha$ and H$\beta$ are in absorption. In the 1~MK low-energy model, both spectra are in emission during the heating phase. The H$\alpha$ line is characterised by a redshifted component with multiple peaks, while the redshifted H$\beta$ peak is severely broadened. In the post-heating phase, both lines are back in absorption, where their cores are Doppler shifted to approximately $+15$~km~s$^{-1}$. In the 3~MK loop, both lines are in absorption during the heating and post-heating phases, showing no significant effects as a consequence to the nanoflare event. 

The intermediate- and high-energy models (both 1 and 3~MK loops) show similar effects in the H$\alpha$ and H$\beta$ lines. During the heating phase, the spectra are double peaked. In the 10~keV models, the blue peak is dominating the spectra in the empty loop, while the red peak dominates the spectra in the dense loop. In the 15~keV dense loop model, the red peak of both lines is more pronounced. The latter is also true for the H$\alpha$ line in the 1~MK loop, while H$\beta$ is almost symmetric around the line core. All heating models show that the line cores are in absorption during the post-heating phase, but some of the line cores are either redshifted or blueshifted as a response to the non-thermal electrons.

\subsection{Formation of line intensity}

\begin{figure*}[ht]
    \centering
    % subfigures without label and numbering (added to the actual figure)
    \subfigure{\includegraphics[width=\columnwidth]{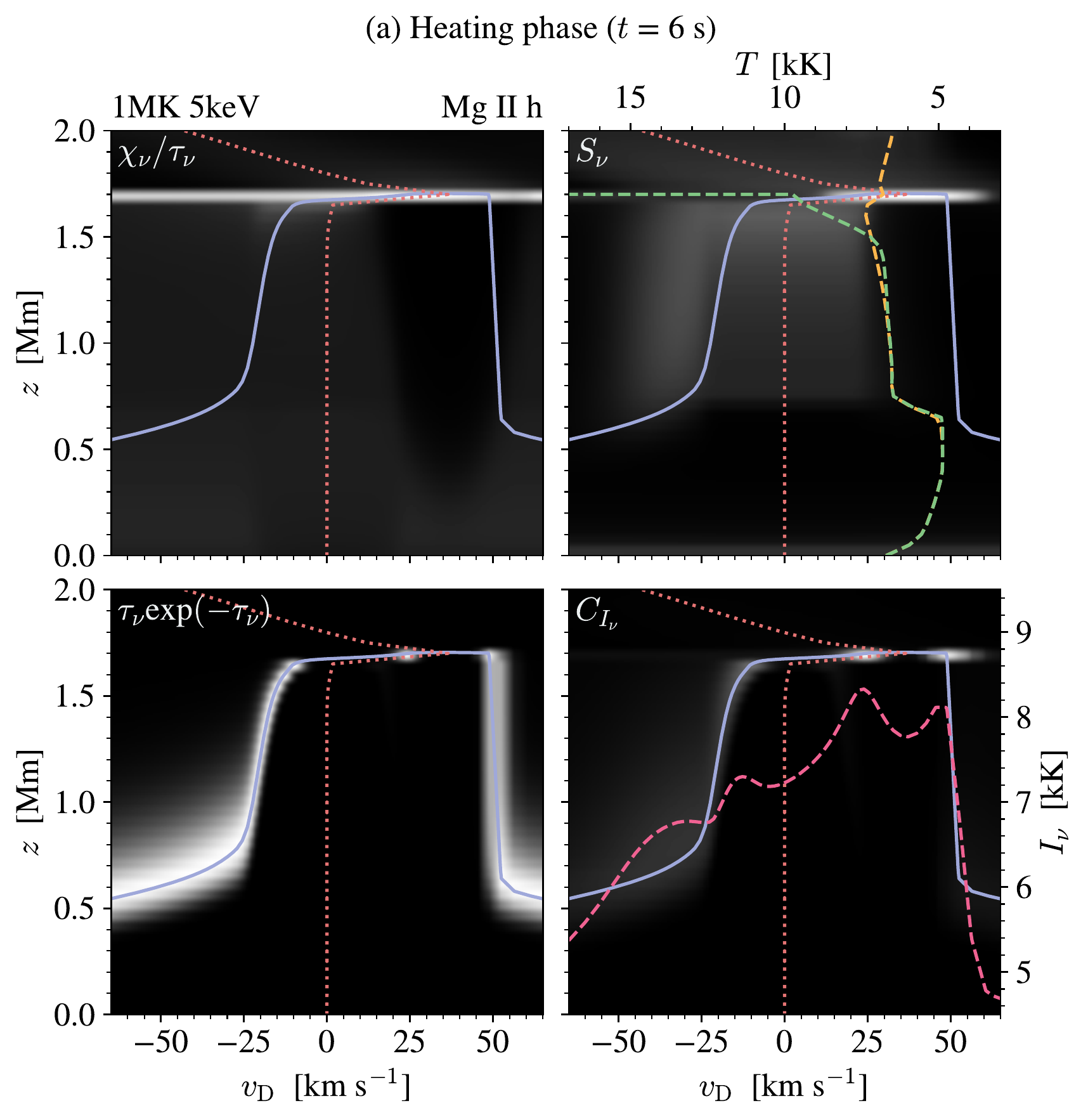}}
    \subfigure{\includegraphics[width=\columnwidth]{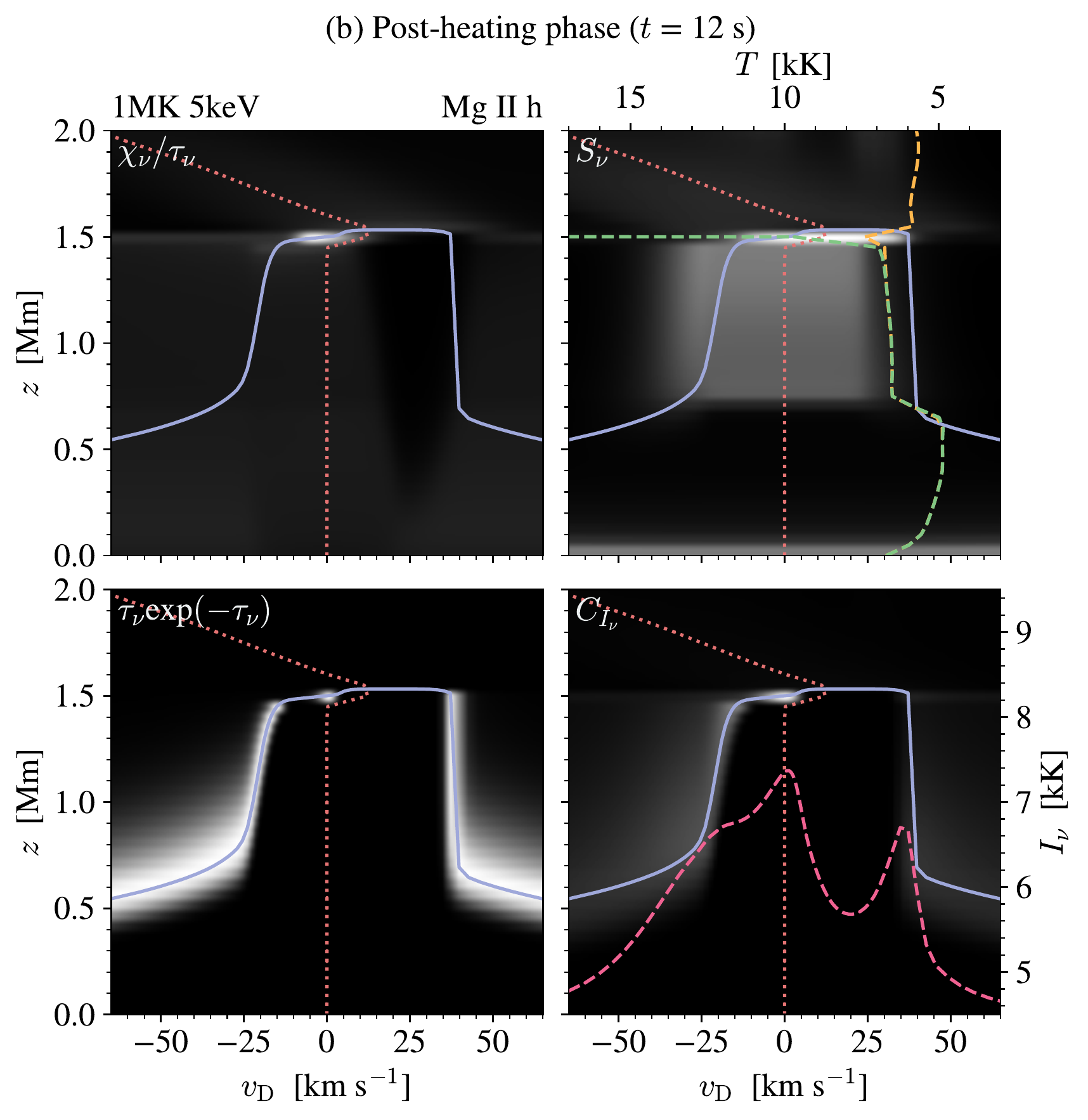}}
    \subfigure{\includegraphics[width=\columnwidth]{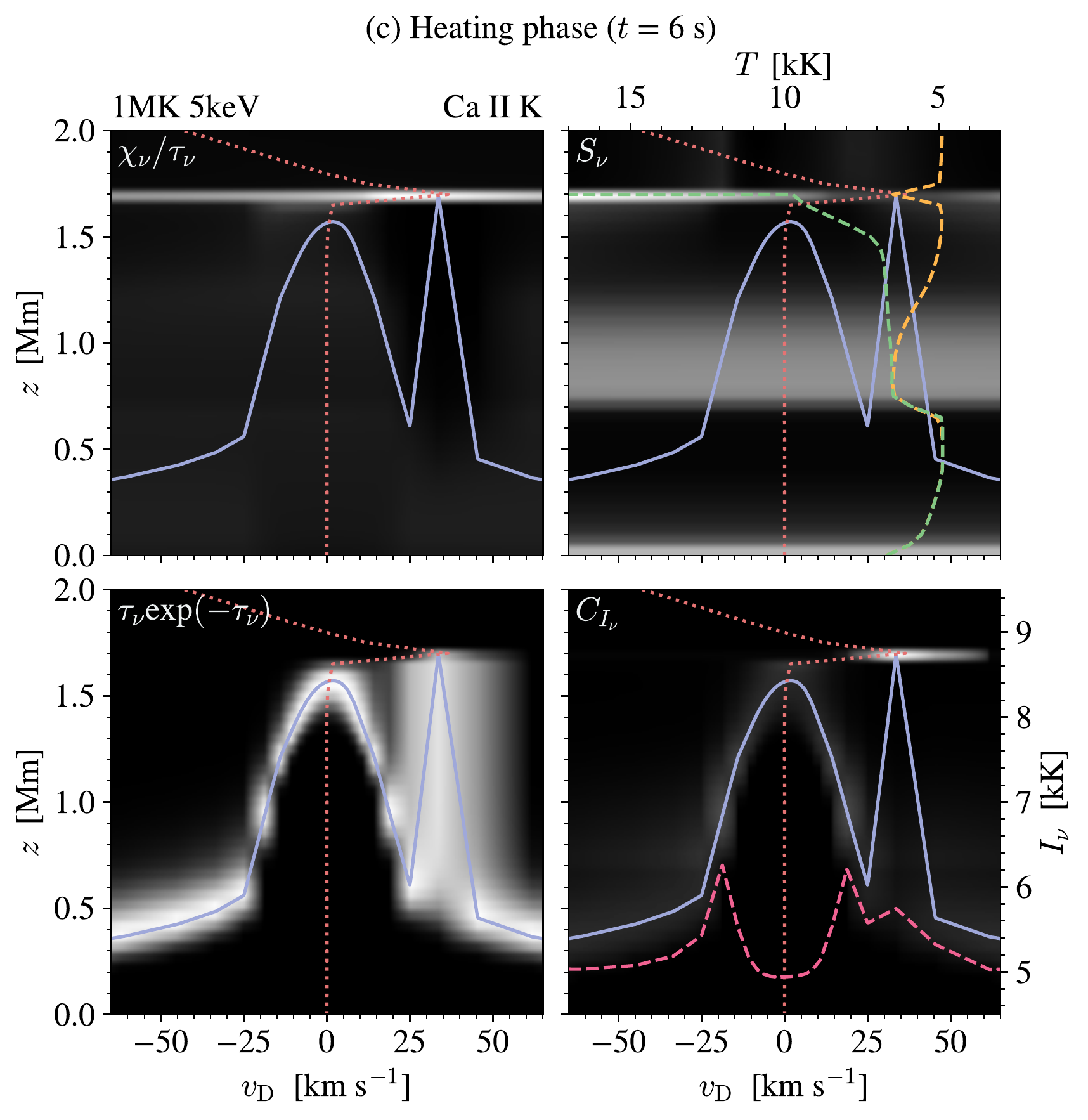}}
    \subfigure{\includegraphics[width=\columnwidth]{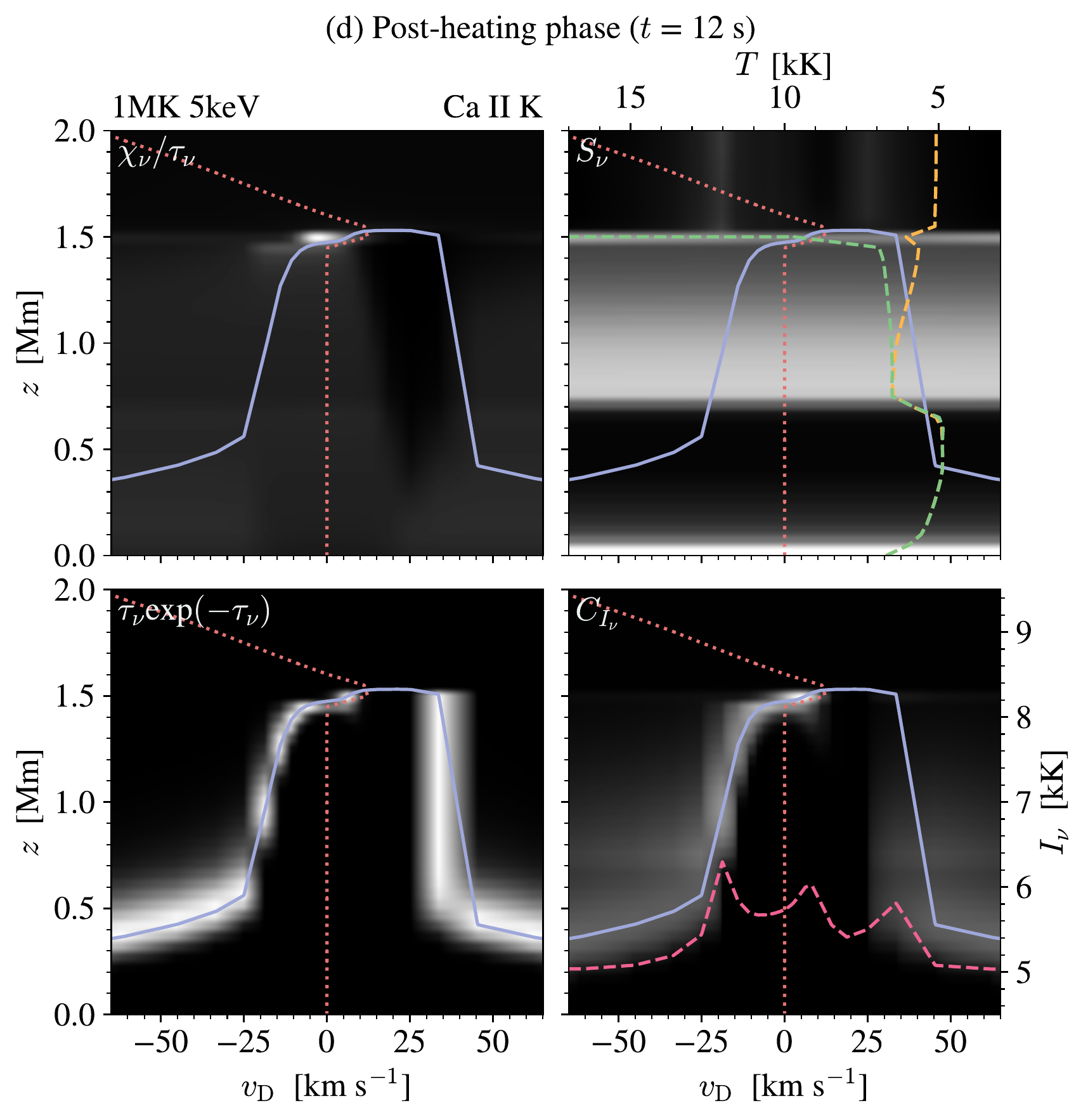}}
    \caption{Intensity formation of the \ion{Mg}{ii}~h (top) and \ion{Ca}{ii}~K (bottom) spectral lines from the 5~keV empty loop model. The subfigures show the heating phase ((a) and (c)) and early post-heating phase ((b) and (d)) for the two lines. The quantities given at the top left corner are shown in greyscale as functions of frequency from line centre (in units of Doppler offset) and height $z$. The $\tau_\nu = 1$ height (purple) and vertical velocity (red dotted) are displayed in all panels. Negative (positive) velocities correspond to upflows (downflows). In addition, the top right panels show the total source function at $v_\mathrm{D} = 0$ (yellow dashed) and Planck function (green dashed) in units of brightness temperature, with high values to the left. The bottom right panels also contain the intensity profile (pink) as brightness temperature. Gamma correction is added to the $\chi_\nu/\tau_\nu$ and $C_{I_\nu}$ terms to amplify the weaker signals.} 
    \label{fig:4p1MK5keV} 
\end{figure*}

\begin{figure*}[ht]
    \centering
    % subfigures without label and numbering (added to the actual figure)
    \subfigure{\includegraphics[width=\columnwidth]{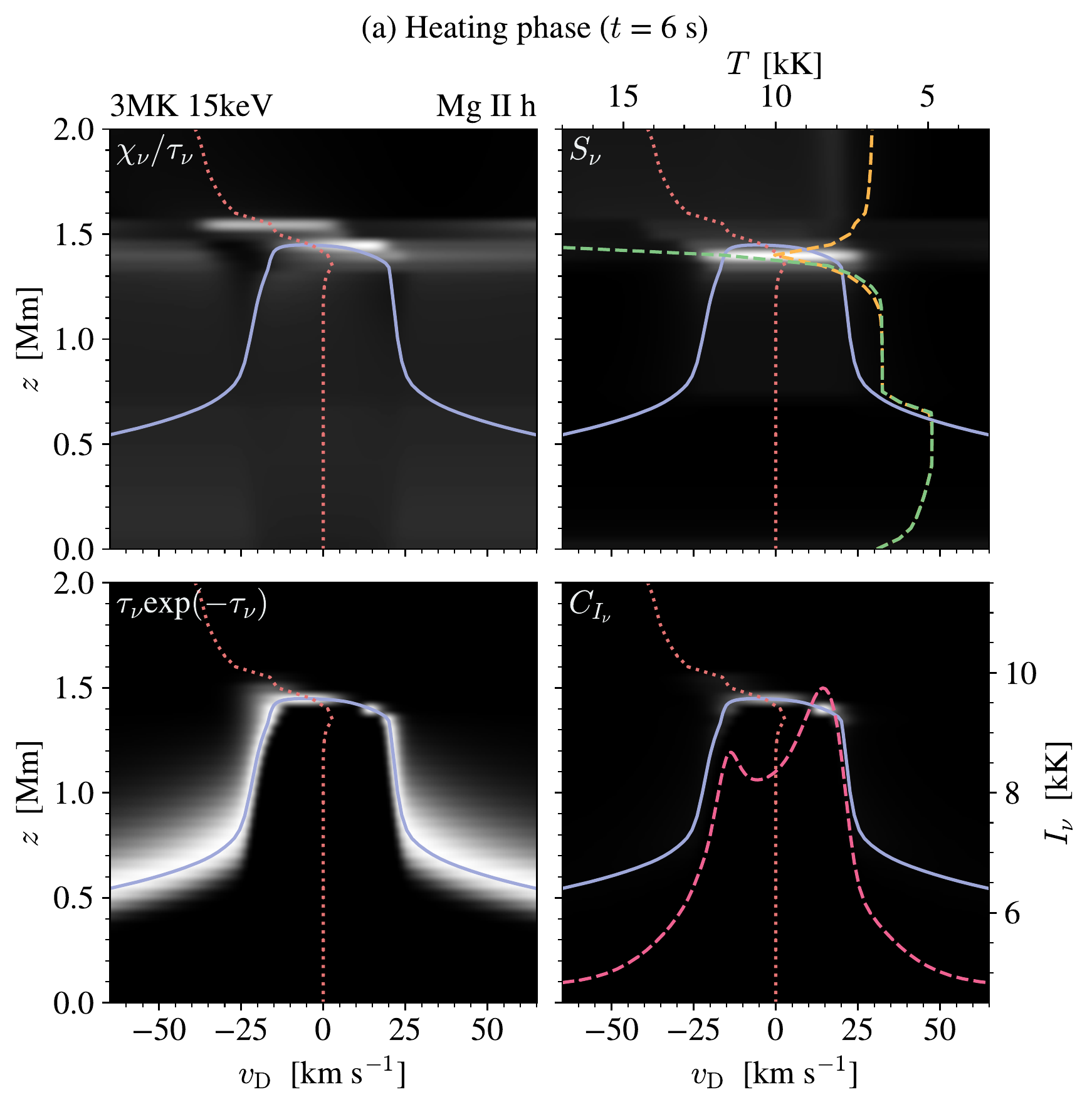}}
    \subfigure{\includegraphics[width=\columnwidth]{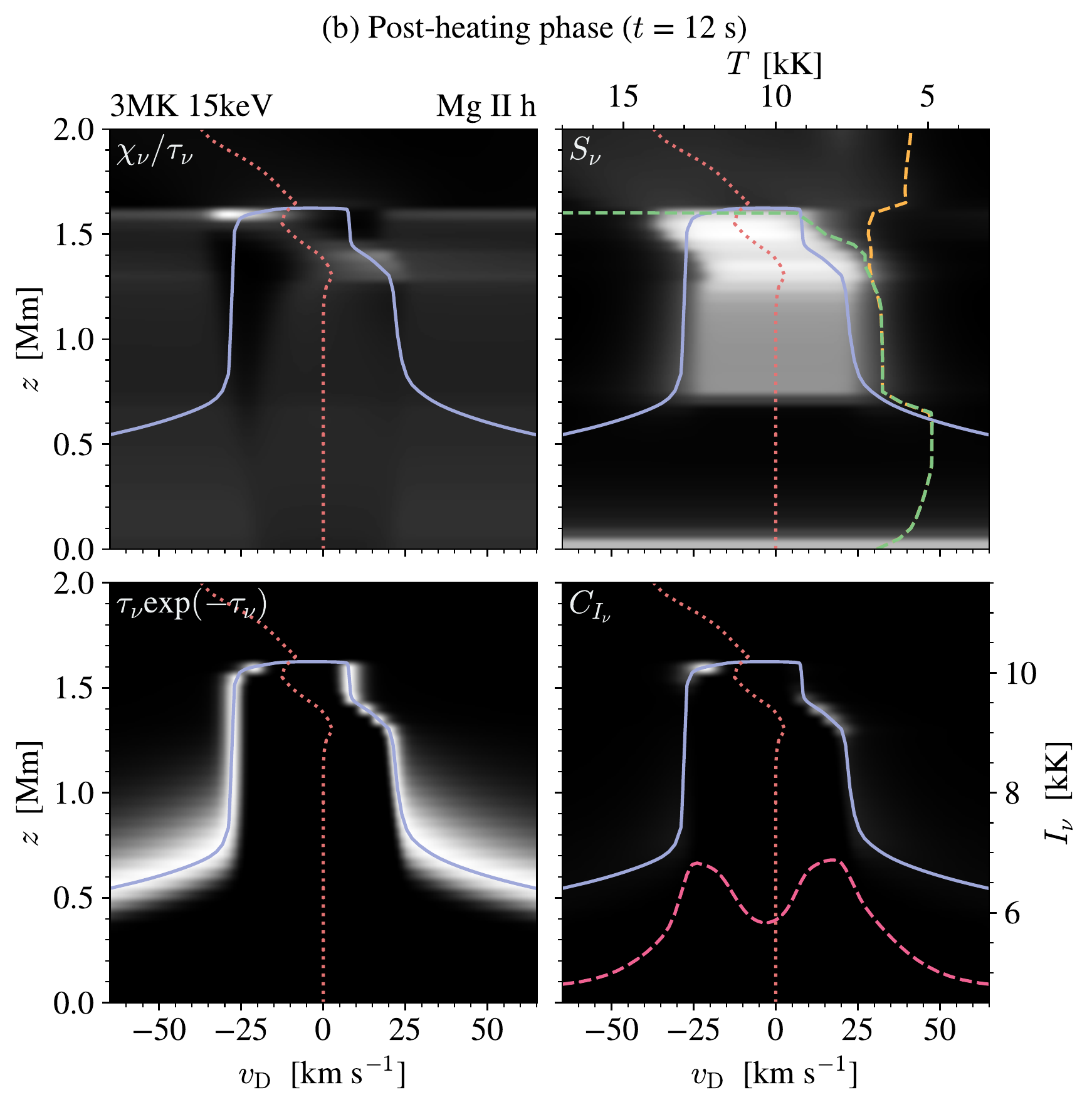}}
    \subfigure{\includegraphics[width=\columnwidth]{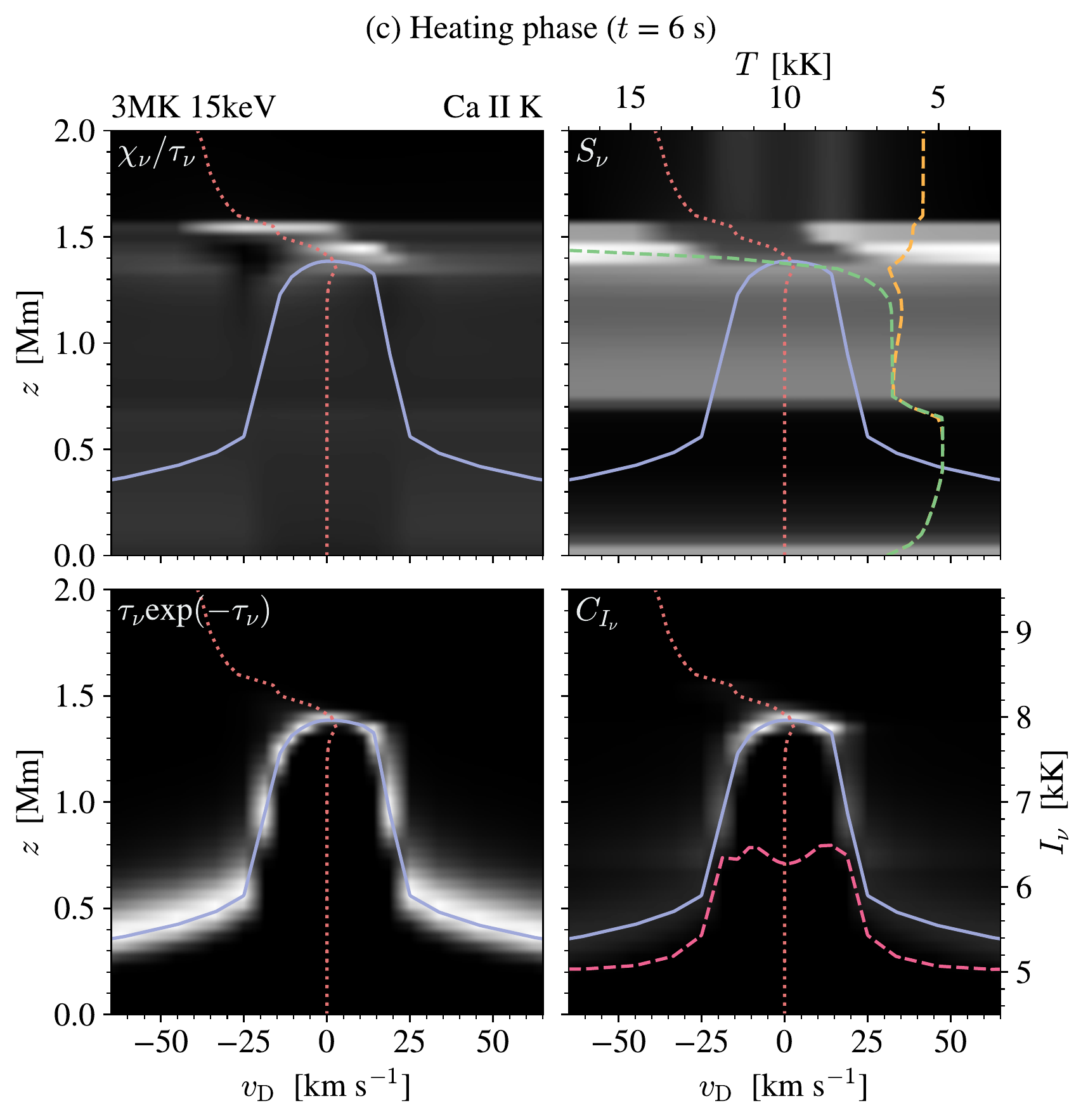}}
    \subfigure{\includegraphics[width=\columnwidth]{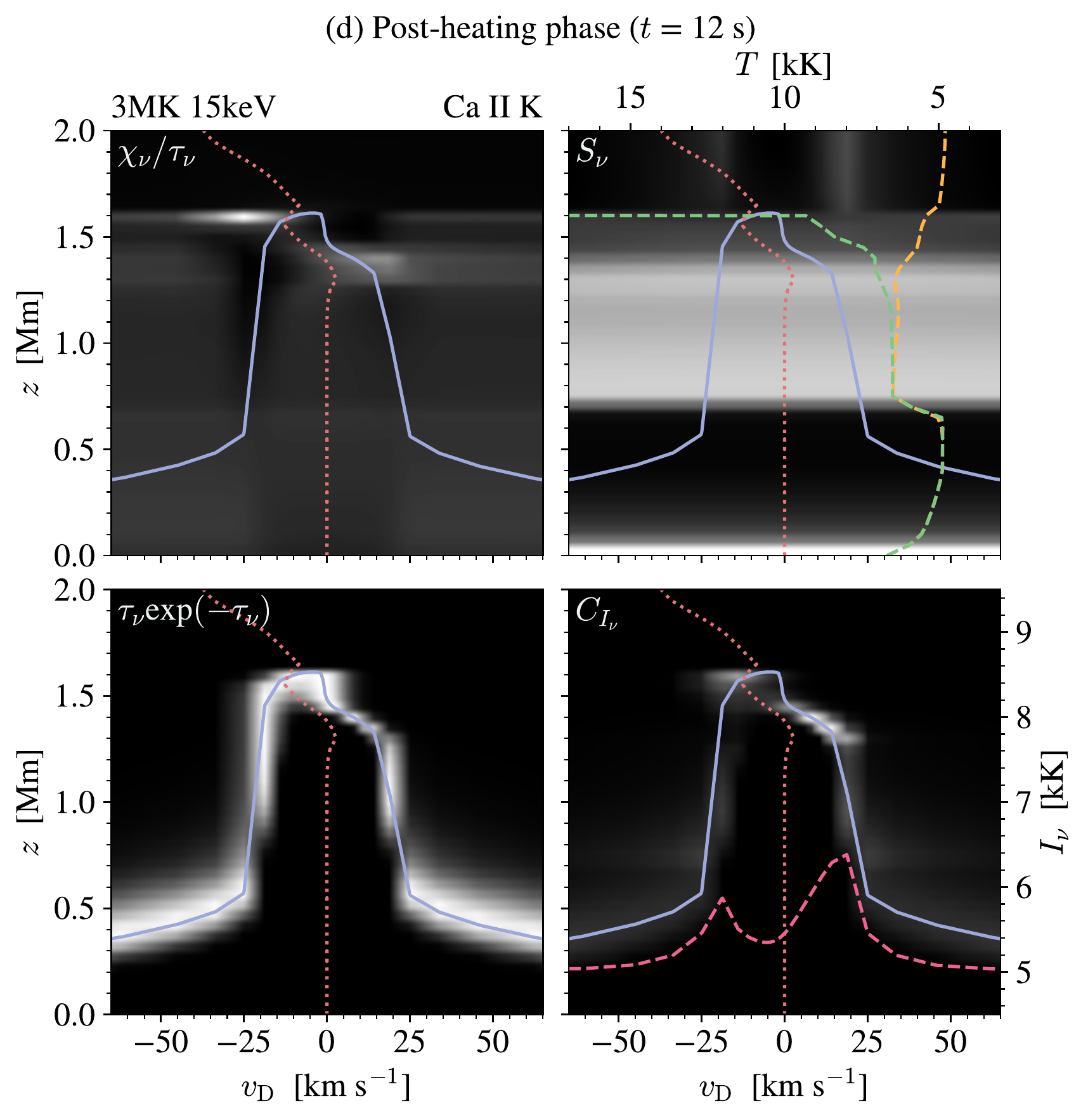}}
    \caption{Intensity formation of the \ion{Mg}{ii}~h (top) and \ion{Ca}{ii}~K (bottom) spectral lines from the 15~keV dense loop model. We note the different scaling of the brightness temperature in the top and bottom panels. See the caption of Fig.~\ref{fig:4p1MK5keV} for more details.} 
    \label{fig:4p3MK15keV} 
\end{figure*}

Figures~\ref{fig:4p1MK5keV} and \ref{fig:4p3MK15keV} show four $2 \times 2$ diagrams of the intensity formation of \ion{Mg}{ii}~h and \ion{Ca}{ii}~K from two of the heating models presented in this work. Because of the extensive parameter survey, we selected two distinctively different models. We present the \ion{Mg}{ii}~h and \ion{Ca}{ii}~K spectra subject to low-energy (5~keV) electrons in the 1~MK loop and high-energy (15~keV) electrons in the 3~MK loop. 
The specifics of the simulations are given in the top left corner of each diagram. The panels in each subfigure represent the terms in Eq.~\ref{eq:cont_func}, and are strongly inspired by the four-panel diagrams in \citet{1997ApJ...481..500C}. % Carlsson&Stein formation

Figure~\ref{fig:4p1MK5keV} (a) shows the \ion{Mg}{ii}~h line during the heating phase in the 5~keV empty loop model. The downward velocity gradient is reflected in the $\chi_\nu/\tau_\nu$ term, exhibiting a strong variation in wavelength. As a result, the $\tau_\nu = 1$ height of the red peak is shifted to larger Doppler offsets. The source function is equal to the Planck function from the photosphere up to around 1.2~Mm where it decouples. Since the Planck function is given in units of brightness temperature, it also represents the temperature itself. The strong rise in temperature at 1.7~Mm gives rise to a significant increase in the source function. The peak in $S_\nu$ is further responsible for the multiple peaks observed in the redshifted component of \ion{Mg}{ii}~h. Investigation of the $\tau_\nu = 1$ height reveals that what appeared to be a double peaked blue component is emission of the blue wing. This is further supported by the increase in $S_\nu$ around the blue wing. The contribution function is strongly dominated by the downward velocity gradient because of its dependence on $\chi_\nu/\tau_\nu$. Signatures of the upflows causing emission in the blue wing are therefore difficult to enhance visually. We have solved this by applying gamma correction to $\chi_\nu/\tau_\nu$ and $C_{I_\nu}$, simply allowing us to highlight the darker regions. 

The intensity formation of \ion{Mg}{ii}~h during the post-heating phase is given in Fig.~\ref{fig:4p1MK5keV} (b). The velocity gradient from the downflowing plasma is weaker compared to the heating phase, and has less impact on the $\chi_\nu/\tau_\nu$ term. The source function is coupled to the Planck function until approximately $z = 1.4$~Mm. The sudden increase in temperature at 1.5~Mm causes the source function to peak, giving rise to two emission peaks in the intensity profile. The increase in temperature and source function gives rise to an additional peak at approximately 0~km~s$^{-1}$, causing the minimum of the central absorption to shift to the red. 

Figure~\ref{fig:4p1MK5keV} (c) shows the intensity formation of \ion{Ca}{ii}~K during the heating phase. The \ion{Ca}{ii}~K line is symmetric between $\pm 25$~km~s$^{-1}$, as the $\tau_\nu = 1$ curve is not affected by the strong velocity gradient in this range. Here, the shape of the profile is mostly defined by the source function, which is coupled to the Planck function up to 0.9~Mm. At around $+35$~km~s$^{-1}$, the $\tau_\nu = 1$ height peaks as a response to the sudden increase in $S_\nu$, which is caused by the increase in temperature due to the strong downflowing velocity. This gives rise to a separate redshifted component at $+35$~km~s$^{-1}$ in the \ion{Ca}{ii}~K line. We note that PRD affects \ion{Ca}{ii}~K less than \ion{Mg}{ii}~h which becomes apparent in the source function panel in the top right: there is little variation in $S_\nu$ along the $x$-axis for most of the atmosphere and $S_\nu$ appears like a horizontal grey band.

\begin{figure*}[!ht]
    \centering
    \includegraphics[width=\textwidth]{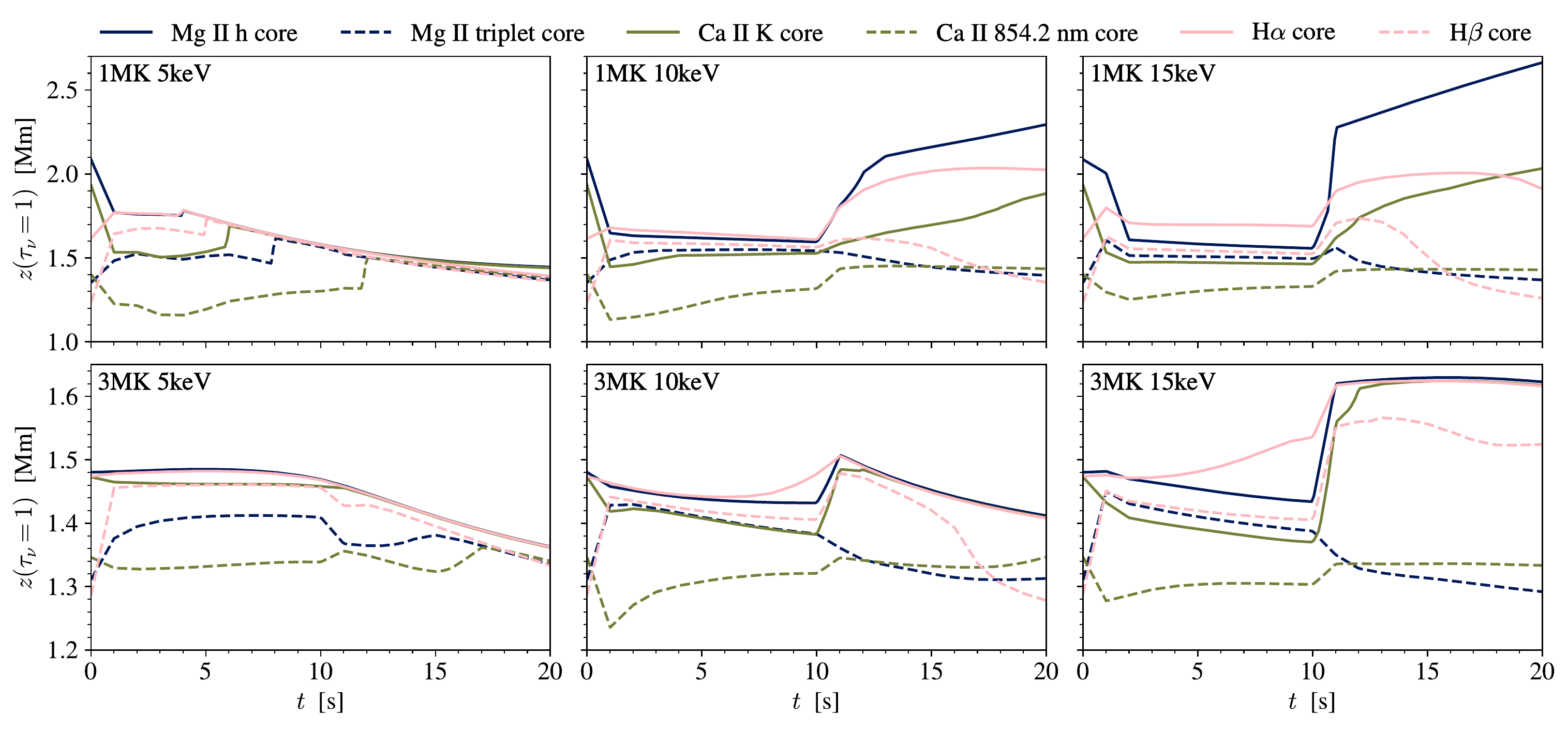}
    \caption{Maximum formation height of the \ion{Mg}{ii}~h, \ion{Mg}{ii}~279.882~nm triplet, \ion{Ca}{ii}~K, \ion{Ca}{ii}~854.2~nm, H$\alpha$, and H$\beta$ line cores. The first 20~s of the simulations are shown, where time is given on the $x$-axis. Each panel represents the different heating models, as stated in the upper left corners.}
    \label{fig:fh}
\end{figure*}

Panel (d) in Fig.~\ref{fig:4p1MK5keV} gives the formation of \ion{Ca}{ii}~K during the post-heating phase. Similar to \ion{Mg}{ii}~h (see panel (b)), the less strong downflowing velocity is reflected in the $\chi_\nu/\tau_\nu$ term, showing the largest increase around 0~km~s$^{-1}$. The sudden increase in temperature at 1.5~Mm causes the source function to peak, giving rise to the peak near the centre of the line at approximately $+7$~km~s$^{-1}$. 
The intensity of the peak at $-20$~km~s$^{-1}$ is stronger than the peak at $+35$~km~s$^{-1}$ as a consequence of the formation of the peak near the line centre. We also note that the contribution function peaks at different Doppler offsets for \ion{Mg}{ii}~h (0~km~s$^{-1}$) and \ion{Ca}{ii}~K ($+7$~km~s$^{-1}$), even though the lines form at the same height. This shows that the velocity is not the only contributing factor to the formation of spectral features, and this underlines the risk of simply interpreting the Doppler offset of spectral features as a direct measure of plasma flows in the atmosphere. 

Figure~\ref{fig:4p3MK15keV} (a) shows the formation of \ion{Mg}{ii}~h during the heating phase in the 15~keV dense loop model. The source function is coupled to the Planck function up until 1.2~Mm, but continues to follow it closely up to around 1.4~Mm. The peak in $S_\nu$ gives a larger contribution to $C_{I_\nu}$ around the red component. As a consequence, the red peak experiences a larger intensity increase and the minimum of the central absorption is shifted to the blue. 

The intensity formation of \ion{Mg}{ii}~h during the post-heating phase is shown in Fig.~\ref{fig:4p3MK15keV} (b). The asymmetric $\tau_\nu = 1$ curve is caused by the variation in velocity between $z = $~1.3--1.6~Mm. The source function is coupled to the Planck function up to 1.3~Mm. From this height, the source function follows a flat slope before it suddenly decreases at 1.6~Mm. The contribution function shows that the peaks of \ion{Mg}{ii}~h form at different heights. The line profile is slightly shifted to the blue, with the minimum of the central absorption at approximately $-5$~km~s$^{-1}$. The blueshift is caused by the general upflow pushing the TR towards greater heights (see Fig.~\ref{fig:3MK_ar} (c)).

Figure~\ref{fig:4p3MK15keV} (c) shows the intensity formation of \ion{Ca}{ii}~K during the heating phase.  
The double peaks in the blue component are caused by the velocity gradient of the upflowing plasma. The source function peaks as a result of the temperature increase, but the upward motion of plasma dilutes the signal such that $S_\nu$ is weaker around the $\tau_\nu = 1$ height of the line core. Still, the contribution to the line core is large enough to increase the intensity, making the emission peaks less pronounced. 

Figure~\ref{fig:4p3MK15keV} (d) shows the \ion{Ca}{ii}~K line during the post-heating phase. The varying velocity of the fluid causes the asymmetric shape of the optical depth unity curve, where the $\tau_\nu = 1$ height increases up to 1.6~Mm at negative Doppler offsets. Similar to the \ion{Mg}{ii}~h line, \ion{Ca}{ii}~K is blueshifted to around $-5$~km~s$^{-1}$. The source function and Planck function are coupled to approximately 0.9~Mm, but does not differ significantly until around 1.4~Mm. As the Planck function (and temperature) increases, the source function decreases. As a result, the source function makes a stronger contribution to $C_{I_\nu}$ at lower heights. This is reflected in the intensity profile, where the red peak is more intense than the blue peak.

Figure~\ref{fig:fh} shows the formation heights of the \ion{Mg}{ii}~h, \ion{Mg}{ii}~279.88~nm triplet, \ion{Ca}{ii}~K, \ion{Ca}{ii}~854.2~nm, H$\alpha$, and H$\beta$ line cores. We include the \ion{Mg}{ii} triplet to investigate how the formation height of another deep chromospheric line compares to \ion{Ca}{ii}~854.2~nm. Since the lines are formed under optically thick conditions, we define the formation height to be the maximum height of the $\tau_\nu = 1$ surface. During the heating phase in the 1~MK heating models (top panels), the \ion{Mg}{ii}~h and \ion{Ca}{ii} cores are formed at lower heights compared to their initial $\tau_\nu = 1$ height, while the \ion{Mg}{ii} triplet, H$\alpha$, and H$\beta$ cores are formed higher in the chromosphere compared to their initial $\tau_\nu = 1$ height. In the 5~keV model, there is a sudden increase in the height of formation for all line cores. This sudden increase is progressively later for spectral lines with lower formation heights, which is caused by local effects from the receding TR (see Fig.~\ref{fig:1MK_ar} (a)). 
As a response to the moving TR, the line cores form over a narrow region in the cooling phase due to the compression of the chromosphere. The formation heights of the line cores in the 10 and 15~keV models increase at 10~s as a result of the sudden decrease in temperature when the heating is turned off. As the temperature continues to decrease, the $\tau_\nu = 1$ heights approach their pre-heating values. 

The bottom panels of Fig.~\ref{fig:fh} represent the 3~MK dense loop models. Since the loop is pre-heated, the TR is located at lower heights compared to the empty loop. The line cores therefore form at lower heights as well. The 5~keV electrons do not affect \ion{Mg}{ii}~h, \ion{Ca}{ii}~K, \ion{Ca}{ii}~854.2~nm, and H$\alpha$ significantly, hence these line cores form at similar heights during the heating phase.
The chromosphere is affected by the intermediate- and high-energy electrons, but the response is slow since the loop is already heated. As a result, the formation heights of the \ion{Mg}{ii}~h and \ion{Ca}{ii}~K cores do not drop as significantly within the first few seconds as in the empty loop. In the 10~keV model, the formation heights decrease as the TR recedes to lower depths. In the 15~keV model, the increase in $z(\tau_\nu = 1)$ after the heating is turned off is a result of the TR moving to greater heights. 

The increase in formation height during the heating phase for the \ion{Mg}{ii} triplet and H$\beta$ cores shows that these lines are sensitive to heating. This result was also found by \citet{2020ApJ...889..124T} % Testa+ variability 
for the \ion{Mg}{ii} triplet. The 5~keV dense loop model even demonstrates that minor atmospheric responses from energy deposited by the non-thermal electrons affect their $\tau_\nu = 1$ heights. In addition, the panels show that \ion{Ca}{ii}~854.2~nm is the deepest diagnostic when electrons are being injected into the loop.

\section{Discussion}

By extending the spectral line diagnostics of \citet{2018ApJ...856..178P} % Polito+ response 
and \cite{2020ApJ...889..124T}, % Testa+ variability
we find that also dominant chromospheric lines in the visible are affected by coronal nanoflare events. The signatures that arise from the non-thermal electron beams highly depend on the initial physical conditions of the loop in combination with the low-energy cutoff $\Ec$ of the power-law distribution of accelerated electrons.
Most importantly, the distance the electrons travel before their energy is deposited is largely dependent on the density of the loop and the energy of the electrons. In low-density loops, there are fewer collisions with the ambient plasma, and vice versa. 
The spectral line signatures are not only affected by the energy deposited directly in the regions where they form, but also by the response to the effects from electron beam energy dissipated elsewhere in the atmosphere, for example plasma flows. 

The most significant effects are seen in the empty loop, where the atmospheric response to $\Ec = 5$~keV electrons causes complex line profiles that are redshifted to large Doppler velocities. Through investigation of the intensity formation, we have shown that the spectral lines are highly affected by the plasma flows through the $\chi_\nu/\tau_\nu$ term. The increase in coronal and TR temperatures leads to the strong downflows responsible for the redshifts, and it is a result of the electrons depositing most of their energy in those regions. 
The large upflows (negative velocities in Fig.~\ref{fig:1MK_ar} (d)) are in agreement with the findings by 
\citet{2014Sci...346B.315T} % Testa+ evidence
and \citet{2015ApJ...808..177R}, % Reep+ optimal 
who showed that low-energy electrons could drive 'explosive evaporation'. 
The atmospheric response to 5~keV electrons strongly depends on the sites where the energy is deposited.

The stopping depth of the electrons in the $\Ec = 10$ and 15~keV energy distributions is deeper into the atmosphere, mainly in the TR and chromosphere. However, we note that the 10~keV electrons in the 3~MK loop deposit more energy in the corona and TR, similarly to the 5~keV models. A portion of the electrons is able to travel to the chromosphere, where the increase in chromospheric emission during the heating phase is significantly larger than in the 5~keV models. The height where the energy is deposited strongly depends on the density of the loop, which becomes apparent when comparing the 10~keV empty and dense loop models. The increased density causes significantly more electrons from the distribution to collide with the ambient plasma at great heights. The atmospheric response to the 10~keV electrons in the dense loop therefore resembles that of the 5~keV models, while the response to intermediate-energy electrons in the empty loop is more similar to the high-energy models. 

A range of energies is represented in the electron energy distributions (see Fig.~\ref{fig:e_dist}), where the input parameters are free and chosen based on observations. The advantage of 1D flare modelling with RADYN is the ability to study specific energetic events in an isolated system. The atmospheric response to the electron beams provides information on features that we can look for in more advanced simulations, such as those produced with the radiative magnetohydrodynamic (MHD) Bifrost code \citep{2011A&A...531A.154G}. % Gudiksen+ bifrost
We previously introduced a Bifrost model including accelerated particles by magnetic reconnection, where the reconnection sites were found by identifying locations where the magnetic topology is not conserved \citep{2018A&A...620L...5B}. % Bakke+ non-thermal
\citet{2020A&A...643A..27F} % Frogner+ accelerated
further developed the modelling of the accelerated particles, and found that the magnetic topology is a significant factor when determining both the amount of accelerated electrons and their energy. In this type of Bifrost models, the magnetic reconnection, and hence particle acceleration, occurs over extended regions. An advantage with such models is the more realistic 3D modelling. However, with the complex configuration of Bifrost models, it is more difficult to control the  parameter space of the simulations.  
RADYN follows the traditional approach of modelling flares (with the injection of electrons at the top of the loop), hence we can more easily isolate the details of the event.
RADYN is also relatively quick to run, which is beneficial when surveying an extensive parameter space to investigate the effects of small-scale events on a variety of spectral lines. 

The signatures in the chromospheric spectra give an indication of what to look for in observations. Complex line profiles as the ones shown in Fig.~\ref{fig:1MK_hm} using the 5~keV empty loop model have not been reported in the literature so far, but might suggest heating by non-thermal electrons if observed. The inclusion of the \ion{Ca}{ii} and \ion{H}{i} lines to the diagnostics gives potential for observing small-scale events with ground-based telescopes, such as the Swedish 1-m Solar Telescope \citep[SST;][]{2003SPIE.4853..341S} % Scharmer+ SST
and the Daniel K. Inouye Solar Telescope \citep[DKIST;][]{2020SoPh..295..172R}. % Rimmele+ 2020 DKIST
With the CHROMIS instrument installed at the SST, it is possible to sample the \ion{Ca}{ii}~H and K and H$\beta$ lines with high spectral resolution. Ground-based telescopes generally allow for higher spatial resolution as compared to millimetre observations and (extreme-)UV diagnostics observed from space. This could possibly provide tighter constraints on the spatial dimensions of the strand widths and temporal variability. We also note that \ion{Mg}{ii}~h and \ion{Ca}{ii}~K appear to be more sensitive to events at the top of the chromosphere than \ion{Ca}{ii}~854.2~nm, and coordinated observations with SST and IRIS, for instance, would be beneficial to provide more constraints on nanoflares. 
A comparison of observations to our numerical results, as well as Bifrost electron beam simulations, will be explored in future work.
\citet{2020ApJ...889..124T} % Testa+ variability
found that \ion{Mg}{ii} profiles from RADYN nanoflare models showed similarities to what was observed at the footpoints of AR coronal loops with IRIS. They also found that the \ion{Mg}{ii} triplet is an important diagnostic for non-thermal electrons due to its sensitivity to heating in the lower atmosphere. We will therefore include the \ion{Mg}{ii} triplet in future spectral line analysis. 

In this paper, we have extended the spectral line diagnostics of \citet{2018ApJ...856..178P} % Polito+ response
and \citet{2020ApJ...889..124T} % Testa+ variability
to include additional chromospheric lines. We have established that non-thermal electrons from small-scale events are able to affect line features through strong plasma flows as a response to the electrons depositing their energy in the corona and TR, but also through energy deposited directly in the chromosphere. We show that the strong chromospheric lines in the visible can also be used as diagnostics of nanoflare-size events.

\begin{acknowledgements}
This research was supported by the Research Council of Norway, project number 250810, % Toppforsk
325491, % ISSRESS
through its Centres of Excellence scheme, project number 262622, and through grants of computing time from the Programme for Supercomputing. P.T. was supported by contract 8100002705 from Lockheed-Martin to SAO, NASA contract NNM07AB07C to the Smithsonian Astrophysical Observatory, and NASA grant 80NSSC20K1272. V.P. and B.D.P. were supported by NASA contract NNG09FA40C (IRIS).
\end{acknowledgements}

\bibliographystyle{aa}
\bibliography{main.bib}

\end{document}